\documentclass[twocolumn,superscriptaddress,floatfix,preprintnumbers]{revtex4}
\usepackage{graphics,amssymb,amsmath,epsfig,color}
\usepackage{graphicx}
\usepackage{subfigure}
\begin{document}

\title{Proposal for the Creation and Optical Detection of Spin Cat States in Bose-Einstein Condensates}
\author{Hon Wai Lau}
\affiliation{Institute for Quantum Science and Technology and Department of Physics and Astronomy, University of Calgary, Calgary T2N 1N4, Alberta, Canada}
\author{Zachary Dutton}
\affiliation{Quantum Information Processing Group, Raytheon BBN Technologies, Cambridge, Massachusetts 02138, USA}
\author{Tian Wang}
\affiliation{Institute for Quantum Science and Technology and Department of Physics and Astronomy, University of Calgary, Calgary T2N 1N4, Alberta, Canada}
\author{Christoph Simon}
\affiliation{Institute for Quantum Science and Technology and Department of Physics and Astronomy, University of Calgary, Calgary T2N 1N4, Alberta, Canada}

\date{\today}

\begin{abstract}
We propose a method to create ``spin cat states'', i.e. macroscopic superpositions of coherent spin states, in Bose-Einstein condensates using the Kerr nonlinearity due to atomic collisions. Based on a detailed study of atom loss, we conclude that cat sizes of hundreds of atoms should be realistic. The existence of the spin cat states can be demonstrated by optical readout. Our analysis also includes the effects of higher-order nonlinearities, atom number fluctuations, and limited readout efficiency.
\end{abstract}

\maketitle


Great efforts are currently made in many areas to bring quantum effects such as superposition and entanglement to the macroscopic level \cite{Monroe,Brune,Leibfried,Lucke,Arndt,Arndt2014,Friedman,Julsgaard,esteve_squeezing_2008,gross_nonlinear_2010,riedel_atom-chip-based_2010,Oconnell,lvovsky_observation_2013,bruno_displacement_2013,Palomaki,Vlastakis}. A particularly dramatic class of macroscopic superposition states are so-called cat states, i.e. superpositions of coherent states where the distance between the two components in phase space can be much greater than their individual size \cite{Monroe,Brune}. For example, the recent experiment of \cite{Vlastakis} created a cat state of over one hundred microwave photons in a waveguide cavity coupled to a superconducting qubit. It was essential for the success of the latter experiment that the loss in that system is extremely small, since even the loss of a single particle from a cat state of this size will lead to almost complete decoherence.

Here we show that it should be possible to create cat states involving the spins of hundreds of atoms in another system where particle losses can be greatly suppressed, namely, Bose-Einstein condensates (BECs), where the spins correspond to different hyperfine states. We use the Kerr nonlinearity due to atomic collisions, which also played a key role in recent demonstrations of atomic spin squeezing \cite{esteve_squeezing_2008,gross_nonlinear_2010,riedel_atom-chip-based_2010}. In contrast to previous proposals \cite{Cirac,Gordon} we do not make use of Josephson couplings to create the cat state, but rely purely on the Kerr nonlinearity in the spirit of the well-known optical proposal of Ref. \cite{Yurke}.

Our approach is inspired by the experiment of Ref. \cite{Zhang}, which stored light in a BEC for over a second. Ref. \cite{Rispe} proposed to use collision-based interactions in this system to implement photon-photon gates, see also Ref. \cite{Vo}. Here we apply a similar approach to the creation and optical detection of spin cat states. Because of the great sensitivity of these states, this requires a careful analysis of atom loss. Our theoretical treatment goes beyond that of Ref. \cite{Rispe}, which was based on the Thomas-Fermi approximation (TFA). Our new approach allows us to study several key imperfections in addition to loss, including higher-order nonlinearities, atom number fluctuations, and inefficient readout, and we conclude that their effects should be manageable.

\begin{figure}
\begin{centering}
\includegraphics[width=0.49\textwidth]{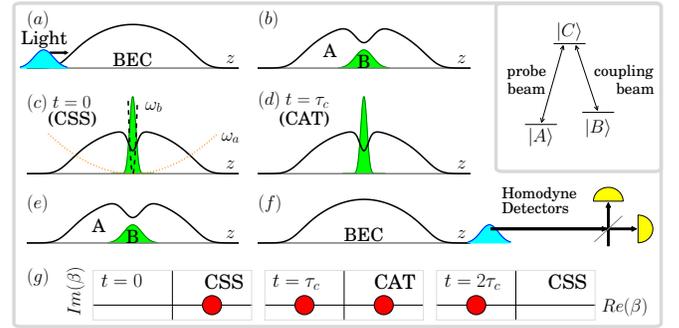}
\par\end{centering}
\caption{\label{cap: scheme-illustration}
(color online) Spin cat state creation (a)-(d)
and detection (e)-(g). In (a)-(f) the radially symmetric photons and spherically symmetric BECs are represented
by spatial density distributions.
(a) A coherent light pulse is sent into the BEC.
(b) The light state is absorbed in the BEC (see inset), creating a CSS. The shape of the input pulse is chosen such that the two-component BEC is in its ground state after the absorption.
(c) The trapping frequency $\omega_{b}$ for the small component is increased adiabatically. The density of the small component now exceeds that of the large component at the center.
(d) The collision-induced Kerr nonlinearity
drives the system into a spin cat state (CAT).
(e) The trapping frequency is adiabatically reduced to its initial value.
(f) The spin state is reconverted into light, whose Husimi $Q$ function \cite{gerry_introductory_2005} is determined via homodyne detection \cite{lvovsky_continuous-variable_2009}.
(g) Expected shape of the $Q(\beta)$ function in phase space.
The coherent state at $t=0$ gives a single peak, while the cat state at $t=\tau_{c}$ yields two peaks.
Further evolution for another interval $\tau_{c}$ returns the output light to a coherent state, yielding a single peak at $t=2\tau_c$. This would not be possible if the two peaks at $t=\tau_c$ corresponded to an incoherent mixture, thus proving the existence of a coherent superposition of CSSs in the BEC at $\tau_c$ \cite{dalvit_decoherence_2000}.
}
\end{figure}


Our scheme is illustrated in Fig. \ref{cap: scheme-illustration}.
The setup is similar to the experiment of Ref. \cite{Zhang}; See also Ref. \cite{lettner}.
In particular, the light is converted into atomic coherences using a control beam ('slow' and 'stopped' light)  \cite{dutton_storing_2004,
Zhang,liu_observation_2001,hau_light_1999,ginsberg_coherent_2007,fleischhauer,phillips}.
We start with a ground state BEC with $N$ atoms in internal states $|A\rangle$.
To create a spin state, a coherent light pulse,
$\left|\alpha\right\rangle_{L}=\sum_{n}c_{n}\left|n\right\rangle $
with mean photon number $\bar{n}=|\alpha|^{2}$ and
$c_{n}=e^{-|\alpha|^{2}/2}(\alpha^{n}/\sqrt{n!})$,
is sent into the BEC (see Fig. \ref{cap: scheme-illustration}a).
The light is absorbed by the BEC and some atoms are converted
into internal states $|B\rangle$ as:
\begin{equation}
  \sum_{n}c_{n}\left|n\right\rangle_L|N,0\rangle_{S}\to|0\rangle_{L}\sum_{n}c_{n}|N-n,n\rangle_{S}:=|0\rangle_L|\alpha\rangle_S, \label{eq: light-to-spin-state}
\end{equation}
where the Fock state $|N_a,N_b\rangle_{S}$ represents $N_a$ and $N_b$ excitations
of wavefunctions $\psi_a$ and $\psi_b$ in the $A$ and $B$ components respectively.
Note that $|\alpha\rangle_S$ is an excellent approximation of a CSS \cite{radcliffe}
$\sum_{n=0}^{N}\sqrt{N!/(n!(N-n)!)}\alpha^{n}|N-n,n\rangle$
in the limit of $N\gg \bar{n}$ which is the case in this scheme.

The described absorption process should prepare the two-component BEC in its motional ground state to avoid the complication of unnecessary dynamics such as oscillations. This can be achieved by matching the shape of the input pulse to the ground state of the effective trapping potential for the small component \cite{Rispe}, provided that the effective trap is not too steep. Once the light has been absorbed, the trapping frequency $\omega_b$ is then increased adiabatically independently of $\omega_a$, which can be achieved by combining optical and magnetic trapping. In the regime $\omega_b \gg \omega_a$, a narrow wavefunction $\psi_b$ is formed at the center and its density can exceed the large component A, see Fig. \ref{cap: scheme-illustration}c. This results in strong self-interaction and hence a large Kerr nonlinearity. On the other hand, keeping $\omega_a$ low reduces the unwanted effects due to collision loss involving the large component.

The spin state will now evolve with time according to
\begin{equation}
  \left|\chi(t)\right\rangle _{S}=\sum_{n}c_{n}e^{-iE(N,n)t/\hbar}|N-n,n\rangle_{S}  \label{eq: spin-state-evolution}
\end{equation}
with $\left|\chi(0)\right\rangle _{S}=|\alpha\rangle_{S}$. If the energy takes the Kerr nonlinear form $\hat{\mathcal{H}}=\hbar\eta_2 \hat{n}^2$, then a spin cat state
$ |\chi(\tau_{c})\rangle _{S} = (|\alpha\rangle _{S}+i|-\alpha\rangle _{S})/\sqrt{2}$
is formed at the time $\tau_c=\pi/|2\eta_2|$ in full analogy with the proposal of Ref. \cite{Yurke}. The problem is thus reduced to the computation of the ground state energy $E(N,n)$.


The energy of the system can be calculated by the following
mean-field energy functional $E[\psi_a,\psi_b;N_a,N_b]$:
\begin{equation}
  E=\sum_{i=a,b}N_{i}\left(\mathcal{K}_{i}+\mathcal{V}_{i}+\frac{1}{2}(N_{i}-1)\mathcal{U}_{ii}\right)+N_{a}N_{b}\mathcal{U}_{ab} \label{eq: total-energy}
\end{equation}
where $\mathcal{K}_{i}$, $\mathcal{V}_{i}$, $\mathcal{U}_{ii}$ and  $\mathcal{U}_{ab}$
are the kinetic energy, potential energy, intra- and inter-component
interaction energy respectively, given by
$\mathcal{K}_{i}=\int(\hbar^{2}/2m)|\nabla\psi_{i}|^{2}$,
$\mathcal{V}_{i}=\int V_{i}|\psi_{i}|^{2}$
with spherically symmetric trapping $V_{i}=m\omega_{i}^{2}r^{2}/2$, and
$\mathcal{U}_{ij}=\int U_{ij}|\psi_{i}|^{2}|\psi_{j}|^{2}$
with interaction strength $U_{ij}=4\pi\hbar^{2}a_{ij}/m$.
Here, $\psi_{i}$ are single particle wavefunctions for $i$-th component with normalization
$\int|\psi_{i}|^{2}d^{3}r=1$, $a_{ij}$ are the scattering lengths, and $m$ is the atom mass. The corresponding dynamic
equation governing the system evolution is the Gross-Pitaevskii equation
(GPE) \cite{esry_hartree-fock_1997,pitaevskii_bose-einstein_2003,pethick_bose-einstein_2008}.
With the restriction $N_{a}=N-n$ and $N_{b}=n$ of spin states creation
in Eq. (\ref{eq: light-to-spin-state}), the nonlinearity
in $n$ can be obtained by the expansion of the energy $E(N,n) = \hbar\eta(N,n)$ around $n=0$ as:
\begin{equation}
  \eta(N,n) = \eta_0(N) + \eta_1(N)n + \eta_2(N)n^{2} + \eta_3(N)n^{3} + ... \label{eq: energy-expansion}
\end{equation}
where $\hbar\eta_{0}$ generates a global phase and $\hbar\eta_{1} = -\mu_{a}+\mu_{b}$
with chemical potential $\mu_{i}$ ($i=a,b$) is the energy to remove
one atom from $|A\rangle$ and add one atom to $|B\rangle$.
$\hbar\eta_{1}$ generates a simple rotation in phase space
$|\alpha\rangle \to |\alpha e^{-i\eta_{1}t}\rangle $,
which can be eliminated by a frame rotation.
The term $\eta_2$ is the Kerr nonlinearity.
We obtain these coefficients by fitting the total energy $E(N,n)$ with $n\in [0,200]$ up to fourth orders in Eq. (\ref{eq: energy-expansion}), where the numerical ground state $\psi_i$ of GPE used in Eq. (\ref{eq: total-energy}) is found by the imaginary time method \cite{chiofalo_ground_2000}.
This numerical approach is better than Ref. \cite{Rispe} because we can avoid the problems associated with the TFA of high densities \cite{pu_properties_1998}. Also, the high density for the small component at the center limits the negative effect of quantum fluctuations in the large component \cite{Trail}. The latter are less important than the classical fluctuations in $\eta_k(N)$ due to uncertainty in $N$, whose effects will be discussed below. Moreover, we can now study the effects of higher-order nonlinearities (in particular $\eta_3$ and $\eta_4$).


\begin{figure}
\begin{centering}
\includegraphics[width=0.49\textwidth]{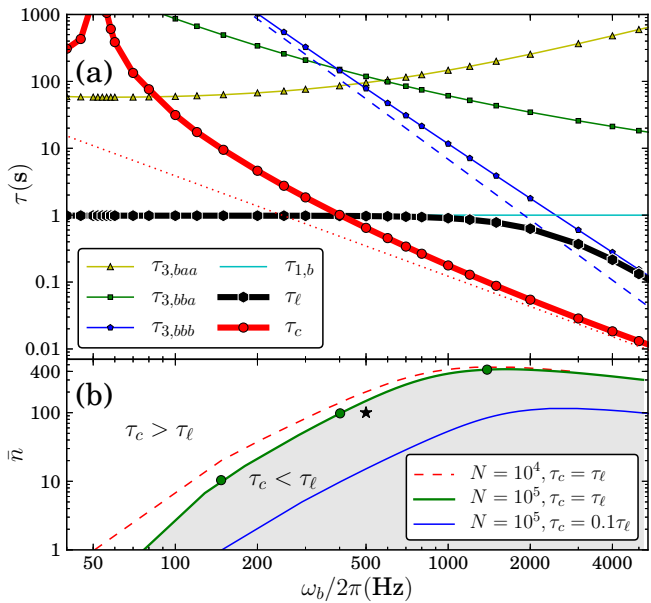}
\par\end{centering}
\caption{\label{cap: Na-loss-phase}
(color online) (a) The time to create a spin cat state  $\tau_{c}=\pi/|2\eta_{2}|$ (thick red curve) versus the time to lose one atom  $\tau_{\ell}$ in component $B$ (thick black curve) as a function of the trapping frequency for the small component $\omega_b$. The size of the cat $\bar{n}=100$ in this example. One sees that $\tau_c < \tau_{\ell}$ is possible for sufficiently large $\omega_b$. The plot also shows the main individual loss channels contributing to the calculation of $\tau_{\ell}$, where $\tau_{m,c}$ is the individual time of losing one atom
through $m$-body collision with particle combinations $c$. It furthermore shows analytic approximations for $\tau_{c}\sim\omega_{b}^{-3/2}$ (red dotted curve) and $\tau_{3,bbb}\sim\omega_{b}^{-3}$ (blue dashed curve), see text.
(b) Achievable cat size $\bar{n}$ as a function of $\omega_b$. The shaded region corresponds to $\tau_{c}<\tau_{\ell}$ for a condensate size $N=10^{5}$ as in (a). The cat size can be increased somewhat by reducing $N$ (dashed line). We also show that there is a region where $\tau_c < 0.1 \tau_{\ell}$ so that loss should really be negligible. The green circles correspond to $\tau_{c}=10,1,0.1 \text{s}$ (from left to right). The star corresponds to the values used in Fig. 3, and the corresponding density distributions are shown in Fig. 1(d). Both plots are for $^{23}$Na with spin states
$|A\rangle = |F=1,m=0\rangle$, $|B\rangle = |F=2,m=-2\rangle$, scattering lengths
$a_{aa}=2.8 \text{nm}$, $a_{bb}=a_{ab}=3.4 \text{nm}$ \cite{Zhang}, loss coefficients
$L_{1}=0.01/ \text{s}$, $L_2=0$, $L_{3}=2\times10^{-42} \text{m}^{6}/\text{s}$ \cite{stamper-kurn_optical_1998}, and a trapping frequency
$\omega_{a}=2\pi\times 20 \text{Hz}$ for the large component.
}
\end{figure}

Fig. \ref{cap: Na-loss-phase} shows our results for the spin cat creation time $\tau_c=\pi/|2\eta_2|$ and achievable cat size $\bar{n}$, taking into account the effects of atom loss. It is clear that the cat time $\tau_{c}$ decreases significantly as the trapping strength $\omega_b$ increases.
Note that the Kerr effect disappears ($\eta_2=0$) around $\omega_b\approx 2\pi\times55$Hz, which may be used for long term storage.
As mentioned above, the reason for the strong Kerr effect for large $\omega_b$ is that strong trapping potential forces $\psi_b$ into a highly localized Gaussian $\phi_{0}(r)=(\frac{m\omega_{b}}{\pi\hbar})^{3/4}e^{-(m\omega_{b}r^2)/2\hbar}$. The radius of $\psi_b$ is of the order
of the characteristic length  $s_b=\sqrt{\pi\hbar/(m\omega_b)}$,
and the density $\rho_b(r)=n|\psi_b(r)|^2$ is peaked at the center
$\rho_b(0)\approx n s_b^{-3}$ which can be much higher than $\rho_a(0)$ in our regime, see Fig. \ref{cap: scheme-illustration}d.
Therefore, the system can be effectively described by
$\hat{\mathcal{H}} \approx \frac{1}{2}U_{bb}\hat{n}(\hat{n}-1)\int d^3r |\phi_0|^4$,
and the second order term is approximately
$\hbar\eta_{2}(N)\approx (U_{bb}/2)\int d^{3}r|\phi_{0}|^{4}=U_{bb}2^{-5/2}s_{b}^{-3}$,
which is consistent with the first order perturbation theory in the Supplemental Materials \cite{supplemental_material}.


The phase between the two components of the spin cat state is flipped by losing just one atom (see the Supplemental Materials \cite{supplemental_material} for more details on the effects of atom loss). This means that $\tau_{c}$ must be smaller than the time to lose one atom $\tau_{\ell}$, which depends on the density and thus $\bar{n}$. In our scheme, the loss of atoms in component $A$ will not affect the cat states directly,
so we focus on the loss of component $B$ only,
which can be estimated by the following loss rate equation \cite{chin_feshbach_2010,li_optimum_2008,li_spin_2009}:
\begin{equation}
  dn/dt = -\tau_\ell^{-1} = -(\mathcal{L}_{1}+\mathcal{L}_{2}+\mathcal{L}_{3})
\end{equation}
where $\tau_{\ell}=1/(\mathcal{L}_{1}+\mathcal{L}_{2}+\mathcal{L}_{3})$ is the approximate time to lose one atom through all possible loss channels if $n\gg1$. The loss rates $\mathcal{L}_{m}$ correspond to the loss through $m$-body collisions involving particles in component B, where $\mathcal{L}_{1}=L_1n$ is due to collisions with the background gas, $\mathcal{L}_{2}=\sum_j L_{2,bj}\int\rho_b\rho_j$ is due to spin exchange collisions, and $\mathcal{L}_{3}=\sum_{j,k} L_{3}\int\rho_b\rho_j\rho_k$ is due to three-body recombination \cite{chin_feshbach_2010}. It is known that the two-particle loss can be eliminated by certain choices of internal states and control methods such as applying a microwave field in \cite{sorensen_many-particle_2001}, or a specific magnetic field as in Ref. \cite{Zhang}. The latter example motivates our choice of parameters in Fig. \ref{cap: Na-loss-phase}.

Fig. \ref{cap: Na-loss-phase}a shows the time to lose one atom through different channels:
$\tau_{1}=(L_{1}n)^{-1}$ for one-body loss and
$\tau_{3,ijk}=(L_{3}\int d^{3}r\rho_{i}\rho_{j}\rho_{k})^{-1}$
for three-body loss with different combination of collisions.
It can be observed that the high $\omega_{b}$ regime is dominated by the loss of $\int \rho_b^3 \sim s_b^{-6} n^3$, which corresponds to $\tau_{3,bbb}$.
For even larger values of $\omega_b$ than those shown in the figure, the three-body loss time $\tau_{3,bbb}$ becomes shorter than $\tau_c$. The small $\omega_{b}$ regime is dominated by the
effect of $\tau_{1}$. See the Supplemental Materials \cite{supplemental_material} for an approximate analytical treatment of atom loss. The desirable region for experiments is $\tau_{c}<\tau_{\ell}$ which also depends on $\bar{n}$.
Therefore, we can draw a $\bar{n}$-$\omega_b$ phase diagram, which shows the achievable cat size as the shaded area in  Fig. \ref{cap: Na-loss-phase}b.


\begin{figure}
\begin{centering}
\includegraphics[width=0.49\textwidth]{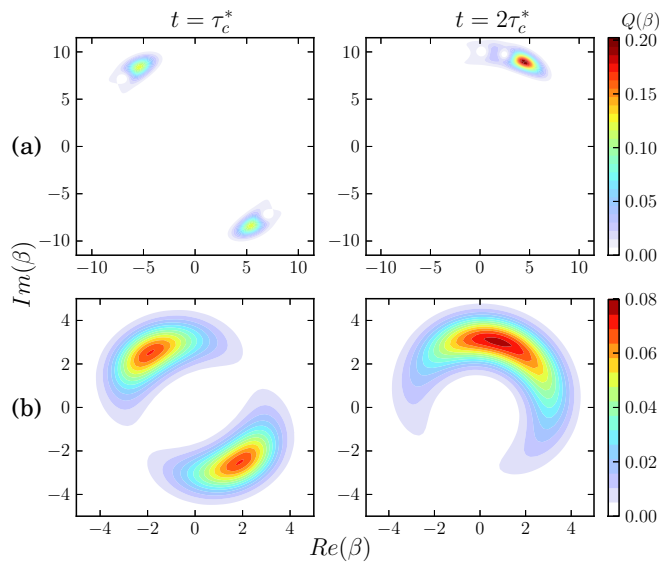}
\par\end{centering}
\caption{\label{cap: Q-results}
(color online) Optical demonstration of the spin cat state in the presence of various imperfections for the parameter values corresponding to the star in Fig. 2(b) ($\bar{n}=100$ and $\omega_b=2\pi\times500$Hz). The spin state is reconverted into light and the Husimi phase space distribution function $Q(\beta)$ is determined via homodyne tomography. (a) Includes the effects of the higher-order nonlinearities $\eta_3$ and $\eta_4$. Two far separated peaks corresponding to the cat state are clearly visible at $\tau_{c}^{*} = 0.68 \text{s}$, and one peak corresponding to the revived coherent state at $2\tau_{c}^{*}$. The shift of the cat creation time due to the higher-order terms is $\tau_c^*/\tau_c=1.06$.
(b) Furthermore includes $90\%$ photon retrieval loss, which moves the peaks towards the origin, and $5\%$ uncertainty in the total atom number, which spreads the peaks in the angular direction.}
\end{figure}

We now discuss how the existence of the spin cat states can be demonstrated via optical readout, see also Fig. 1(e) to 1(g). Our detection scheme is based on a revival argument and hence involves measurements at different times \cite{dalvit_decoherence_2000} (See also the related experiment of Ref. \cite{greiner}). In all cases the readout process starts by reducing the trapping frequency adiabatically to its initial value. Then the spin state $\left|\chi(t)\right\rangle _{S}$ is reconverted into a state of light $\left|\chi(t)\right\rangle _{L}$, followed by homodyne detection on the output light. Using optical homodyne tomography \cite{lvovsky_continuous-variable_2009}, we can reconstruct the Husimi Q-function \cite{gerry_introductory_2005}
$Q(\beta,t)=\frac{1}{\pi}\left\langle \beta\left|\hat{\rho}(t)\right|\beta\right\rangle $
with the density matrix $\hat{\rho}(t)=\left|\chi(t)\right\rangle _{L}\left\langle \chi(t)\right|$.
The Q-function allows us to visualize the resulting spin states of BEC as a function of time.

Higher-order nonlinearities distort the cat state and shift the cat creation time from $\tau_c$ for a pure Kerr nonlinearity to a different observed value $\tau_c^*$.
Fig. \ref{cap: Q-results}(a) shows $Q(\beta,\tau_c^*)$ for $\omega_b=2\pi\times500$Hz including up to fourth-order nonlinear terms $\eta_k$. Two peaks at $t=\tau_{c}^{*}$ can be identified clearly. At the revival time $t=2\tau_{c}^{*}$, a single peak is recovered, which proves the existence of spin cat states in the BEC at $\tau_{c}^{*}$, as described in Fig. \ref{cap: scheme-illustration}(g).
Note that the definition of $\tau_{c}^{*}$ used is the time at which the Q-function shows the two highest peaks.
In general, $\eta_3 < 0$ and hence $\tau_c^*>\tau_c$ for $\omega_b\gg\omega_a$ since $\psi_b$ is less localized than $\phi_0$ due to the repulsive self-interaction.
For the weakly phase separated regime ($a_{aa} a_{bb} \lesssim a_{ab}^2$) used in Fig. \ref{cap: Na-loss-phase}, the effective compression from component A on $\psi_b$ can have the reverse effect. This gives $\eta_{3} \approx 0$ and thus
nearly perfect cat states at $\omega_{b}\approx2\pi\times400$ Hz. Further higher-order effects are shown in the Supplemental Materials \cite{supplemental_material}.

In current experiments the light storage and retrieval process involves significant photon loss, e.g. about 90\% loss in Ref. \cite{Zhang}.
Its main effect is to move the peaks towards the origin, see Fig. 3b and Supplemental Materials \cite{supplemental_material}. One important requirement for achieving high absorption and emission efficiency is high optical depth. For the example of Fig. \ref{cap: Na-loss-phase}, the optical depth can be estimated as  $d\sim N\lambda^{2}/(\pi R^{2})=34$
with $N=10^5$, wavelength $\lambda= 590$nm and the BEC radius $R=18\mu$m. This is in principle sufficient to achieve an overall efficiency close to 1 \cite{gorshkov}.


Another important experimental imperfection is the fact that the total atom number $N$ cannot be precisely controlled from shot to shot. This leads to fluctuations in the nonlinear coefficients $\eta_k$. The most important negative effect of these fluctuations is dephasing, i.e. angular spreading of the peaks in Fig. 3 in phase space \cite{wang}. The magnitude of the angular spread at the time $\tau_{c}=\pi/|2\eta_{2}|$ of the cat state creation can be estimated as $\Delta \varphi = \frac{\pi \Delta N}{2\eta_2} \sum_k k \bar{n}^{k-1} \frac{\partial \eta_k}{\partial N}$, where $\Delta N$ is the uncertainty in $N$, as discussed in more detail in the Supplemental Materials \cite{supplemental_material}. We find that the sensitivity of our scheme to atom number fluctuations is minimized for $\omega_b \approx 2\pi\times 600$Hz. Fig. 3b shows that a 5\% uncertainty in $N$ can be tolerated for $\omega_b=2\pi\times 500$ Hz (even when occurring in combination with 90 \% photon loss).


Two key ingredients for the success of the present scheme are the use of a high trapping frequency for the small component and the achievement of very low loss. The high trapping frequency enhances the strength of the Kerr nonlinearity, making it possible to create cat states without relying on a Feshbach resonance as proposed in Ref. \cite{Rispe}.  This makes it possible to avoid the substantial atom loss typically associated with these resonances \cite{chin_feshbach_2010}, and also allows one to use the magnetic field to eliminate two-body loss, which is critical. For example, the loss rates for the choice of Rubidium internal states discussed in Ref. \cite{li_spin_2009} would only allow cat sizes of order ten atoms, see the Supplemental Materials \cite{supplemental_material}. The high trapping frequency also helped us to suppress the unwanted effects of higher-order nonlinearities and atom number fluctuations. If the readout efficiency could be increased significantly, then the present scheme could also be used to create optical cat states. Besides their fundamental interest, both spin cat states and optical cat states are attractive in the context of quantum metrology \cite{bjork}.


We thank L. Hau, C. Trail, D. Feder, K. Almutairi and B. Sanders for useful discussions. This work was supported by AITF and NSERC.

\pagebreak
\clearpage
\widetext
\begin{center}
\textbf{\large Supplemental Materials: Proposal for the Creation and Optical Detection of Spin Cat States in Bose-Einstein Condensates}
\end{center}
\setcounter{equation}{0}
\setcounter{figure}{0}
\setcounter{table}{0}
\setcounter{page}{1}
\makeatletter
\renewcommand{\theequation}{S\arabic{equation}}
\renewcommand{\thefigure}{S\arabic{figure}}
\renewcommand{\bibnumfmt}[1]{[S#1]}
\renewcommand{\citenumfont}[1]{S#1}

\section{Properties of two-component BEC}

\begin{figure}
    \centering
    \subfigure[Full width at half maximum (FWHM) for density $\rho_b$]{\includegraphics[width=0.45\textwidth]{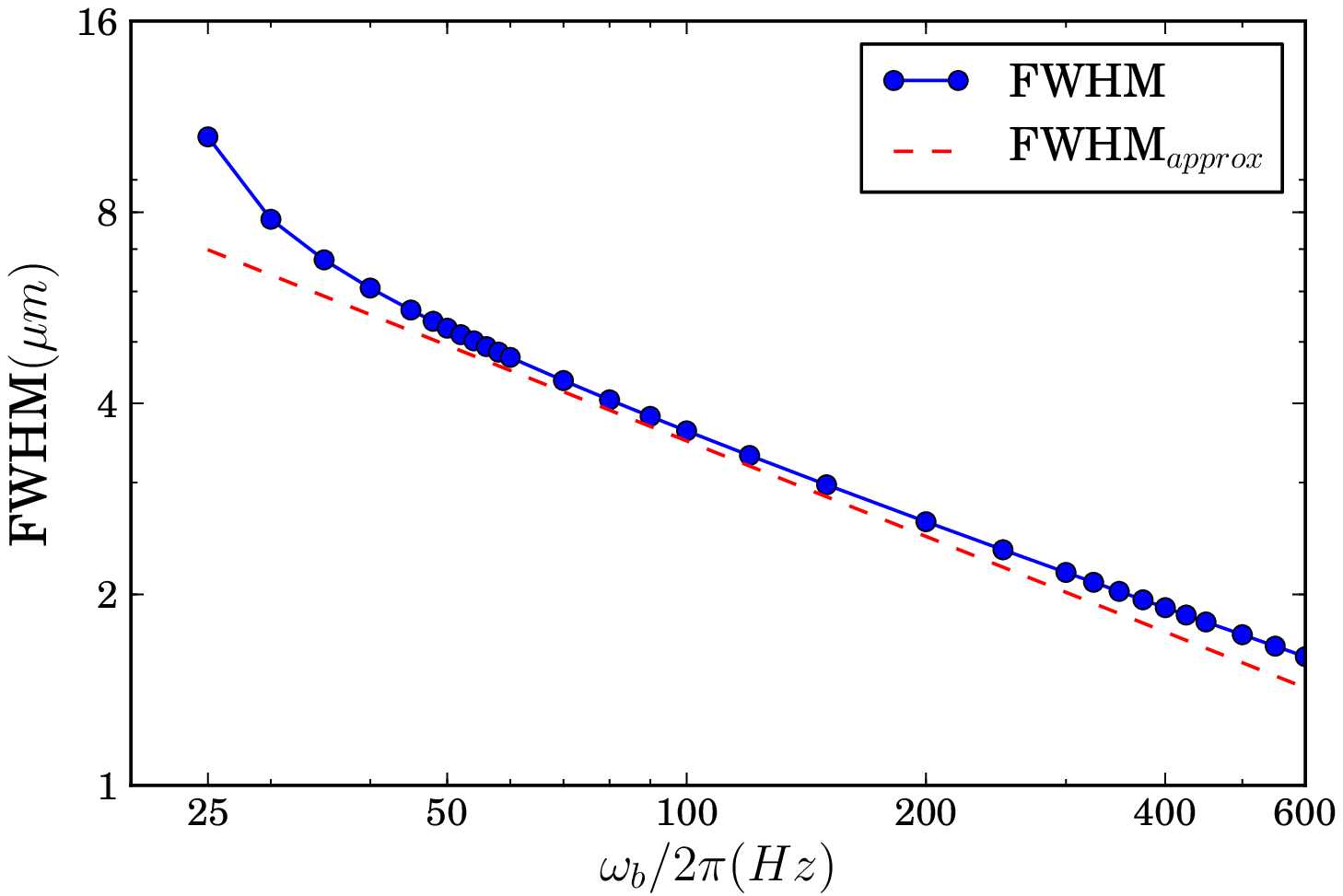}}
    \subfigure[Density distribution at center for both components]{\includegraphics[width=0.45\textwidth]{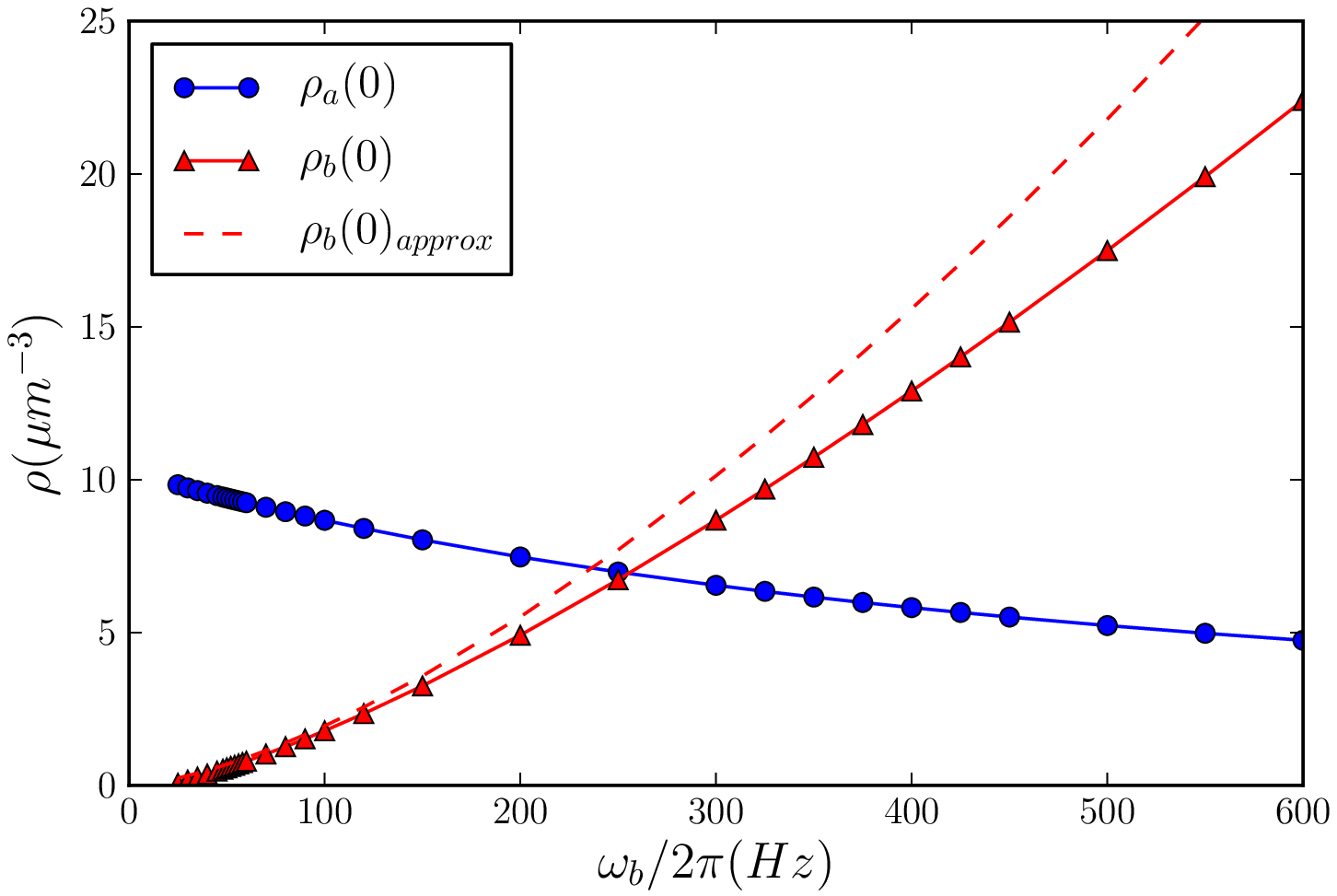}}
    \vfill
    \subfigure[Linear coefficient $\eta_1$ and its approximation Eq. (\ref{eq: eta1-approx})]{\includegraphics[width=0.45\textwidth]{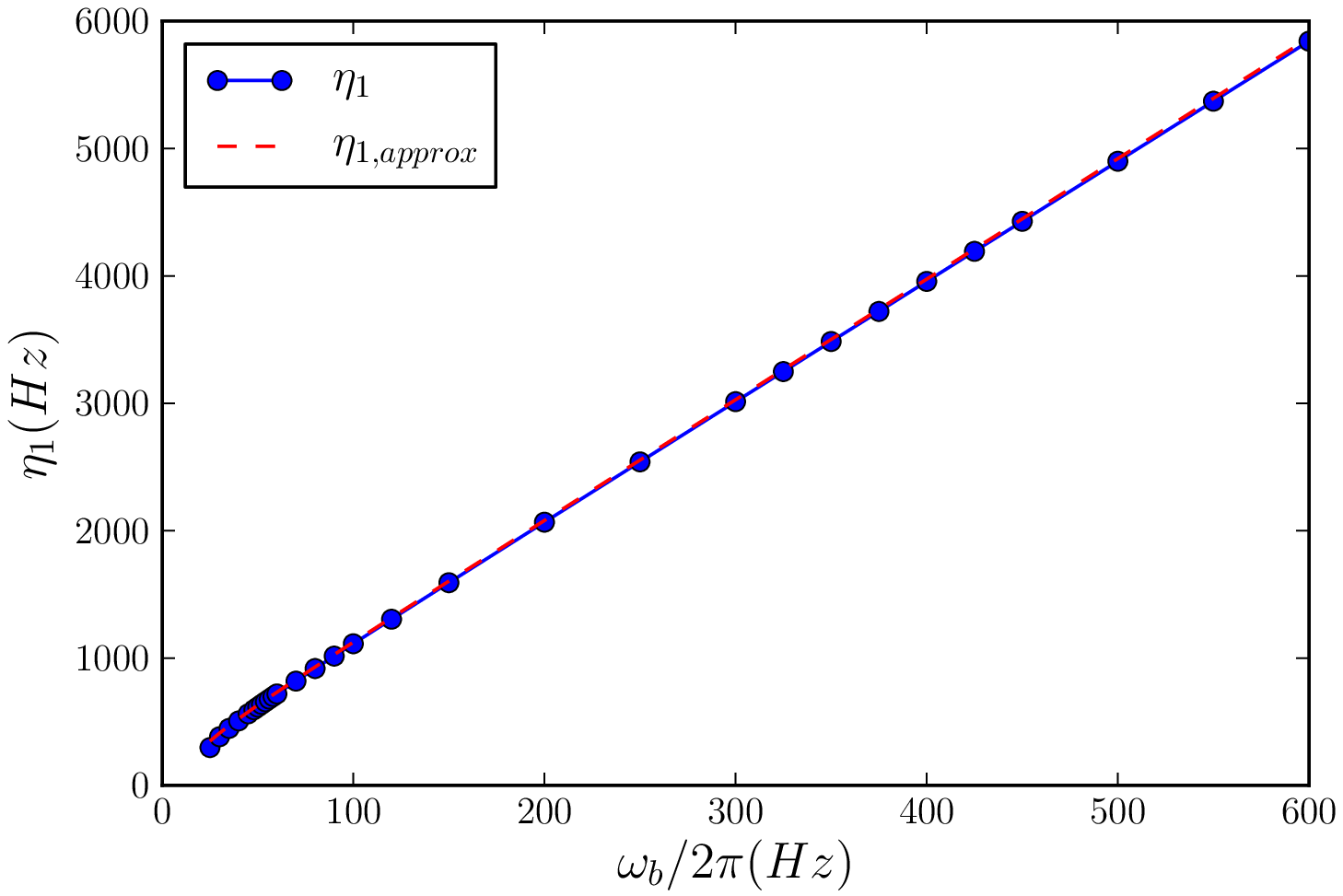}}
    \subfigure[Kerr coefficient $\eta_2$ and its approximation Eq. (\ref{eq: eta2-approx-simple})]{\includegraphics[width=0.45\textwidth]{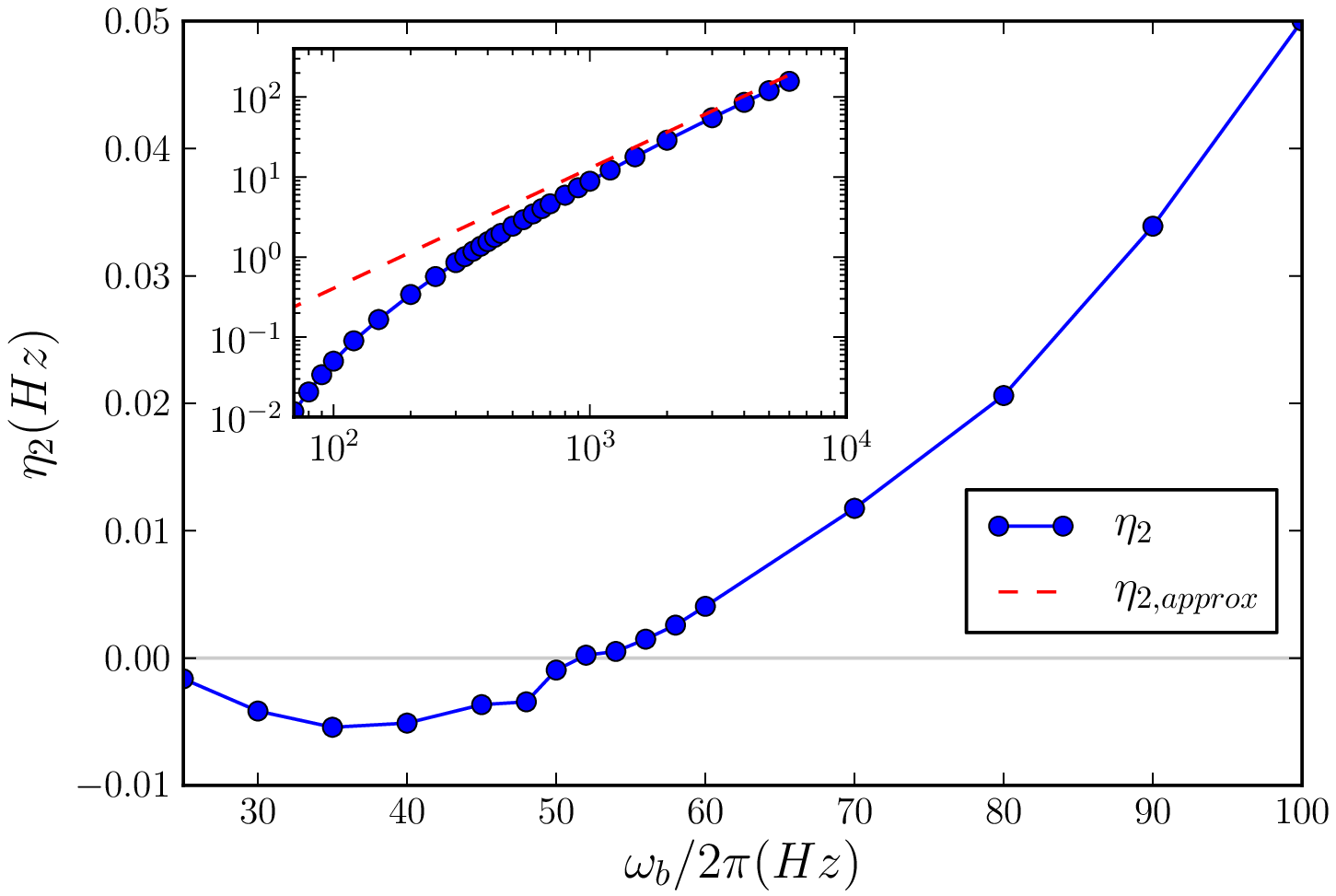}}
    \vfill
    \subfigure[Third order coefficient $\eta_3$]{\includegraphics[width=0.45\textwidth]{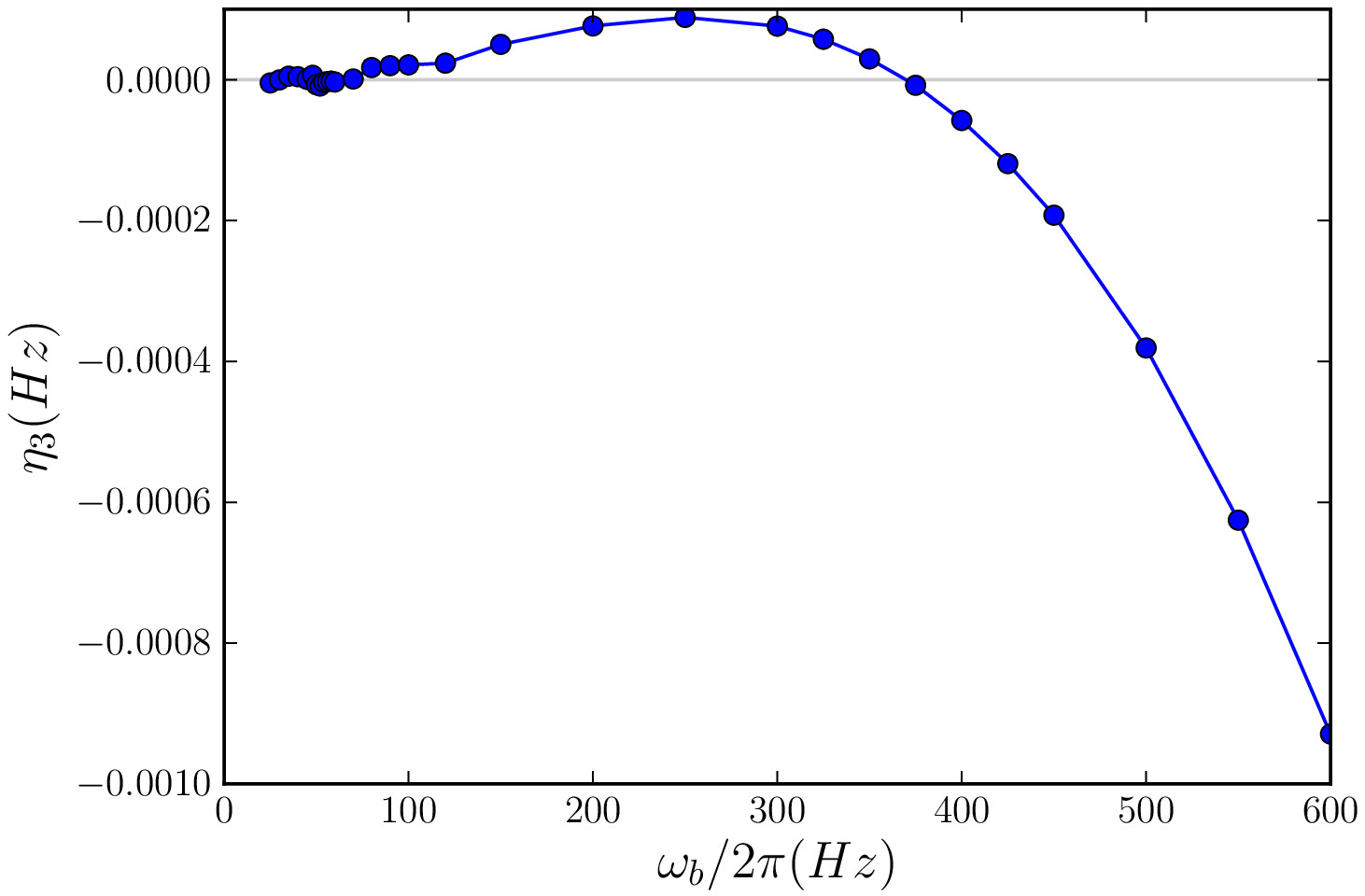}}
    \subfigure[The shift of the `best' cat time $\tau_c^*/\tau_c$]{\includegraphics[width=0.45\textwidth]{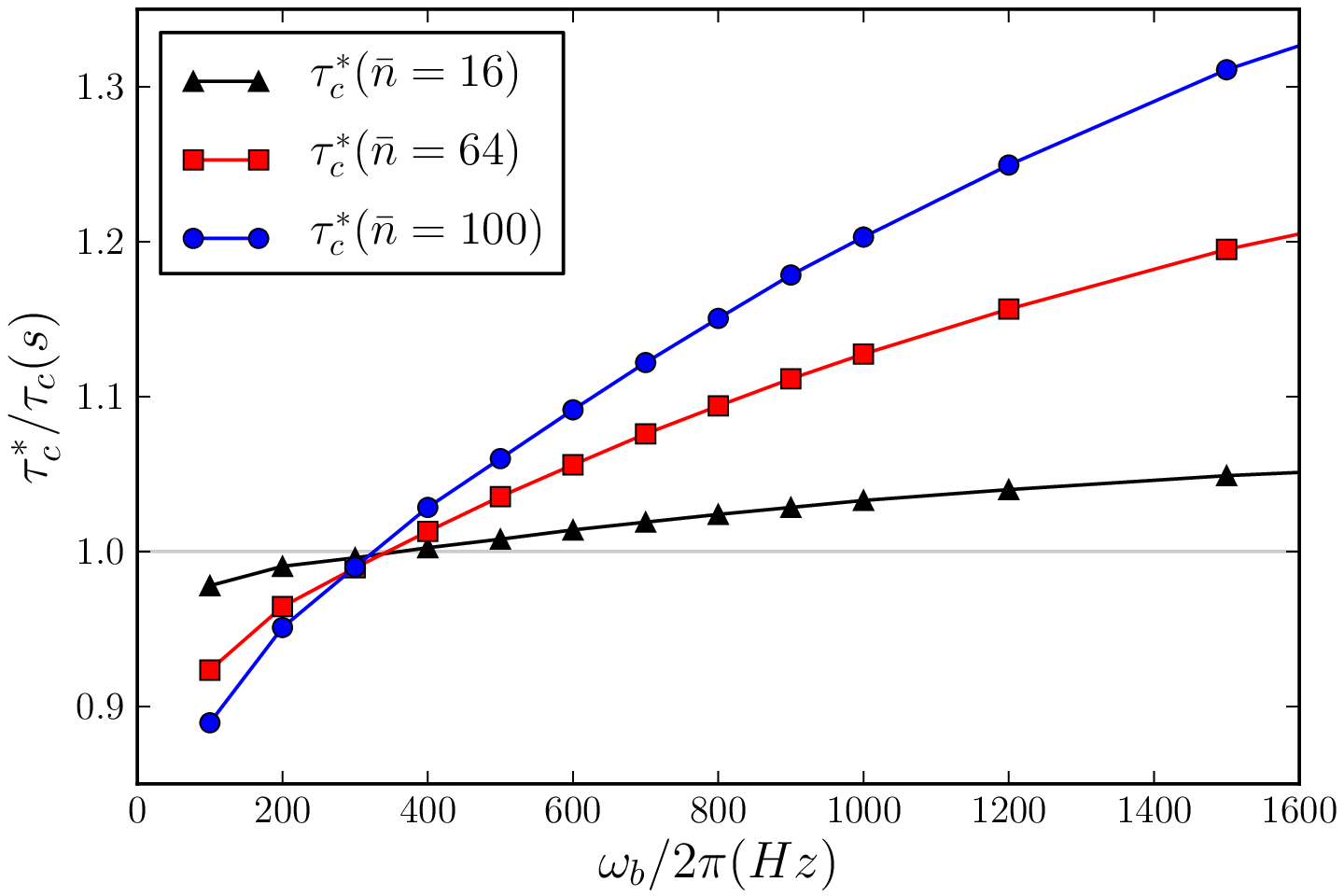}}
\caption{\label{cap: Properties}
Properties of spin states in the two-component BEC for the scheme with cat size $\bar{n}=100$.
(a) The width of component B becomes close to the width of a Gaussian as in Eq. (\ref{eq: GPE-unpert-sol}) around $\omega_b\approx 2\pi50 \mbox{Hz}$ (red dash curve).
The deviation at high $\omega_b$ is because of self-repulsion in component B.
Also, in this weakly phase separated regime $a_{aa} a_{bb} \lesssim a_{ab}^2$ with equal trapping $\omega_a=\omega_b=2\pi20 \mbox{Hz}$, the component B is located outside of component A.
The component B only peaks at the center with $\omega_b$ about $10\%$ higher than $\omega_a$.
(b) Density $\rho_{a}(r=0)$ and $\rho_{b}(r=0)$ at the center of the trap.
Note that the density $\rho_b(0)$ becomes greater than $\rho_a(0)$  around $\omega_{b}/2\pi\approx250 \mbox{Hz}$ (see Fig. \ref{cap: wavefunc-demo} for a spatial distribution).
This suggests that most effects from the main BEC component A, including its quantum depletion, should be relatively small beyond $\omega_{b}/2\pi>250 \mbox{Hz}$.
The red dashed curve is the density of the Gaussian approximation Eq. (\ref{eq: GPE-unpert-sol})
(c) The numerical results for $\eta_{1}$ show a good agreement with  first order perturbation theory.
(d) The numerical solution for $\eta_{2}$ crosses zero around $\omega_b/2\pi\approx 55 \mbox{Hz}$, which causes the cat time $\tau_{c}=\pi/|2\eta_{2}|$ to diverge around this point. The inset shows that the numerical results approach the simple scaling $\eta_{2}\sim\omega_{b}^{3/2}$ at large $\omega_{b}$.
(e) The third order term $\eta_{3}$ also shows a zero-crossing point at around $\omega_{b}/2\pi\approx 375 \mbox{Hz}$, which is a good region to observe nearly perfect cat states with small $\bar{n}$.
(f) The relative change of the best real cat time $\tau_{c}^{*}$ from $\tau_{c}=\pi/|2\eta_2|$. The region with $\tau_{c}^{*}/\tau_{c}>1$ is roughly $\omega_{b}/2\pi \gtrsim 375 \mbox{Hz}$ depending on $\bar{n}$, which corresponds roughly to the region $\eta_{3}<0$ in subfigure e, and vice versa. Note that the fourth order term is included when determining $\tau_{c}^{*}$, see text for its definition and Fig. \ref{cap: Qmax-tauc}.
The parameters used here are the same as in Fig. 2 in the main text: $^{23}\mbox{Na}$ with spin states
$|A\rangle = |F=1,m=0\rangle$, $|B\rangle = |F=2,m=-2\rangle$, scattering lengths
$a_{aa}=2.8 \text{nm}$, $a_{bb}=a_{ab}=3.4 \text{nm}$ \cite{S_Zhang}, loss coefficients
$L_{1}=0.01/ \text{s}$, $L_2=0$, $L_{3}=2\times10^{-42} \text{m}^{6}/\text{s}$ \cite{S_stamper-kurn_optical_1998}, and a trapping frequency
$\omega_{a}=2\pi\times 20 \text{Hz}$ for the large component.
}
\end{figure}

The most important results in the main text and these Supplemental Materials are based on numerical methods.
Therefore, the results can be considered exact within the domain of validity of the equations we used, without relying on analytic approximations.
The two-component BEC can be described by the mean-field Gross-Pitaevskii equation (GPE) \cite{S_esry_hartree-fock_1997,S_pitaevskii_bose-einstein_2003,S_pethick_bose-einstein_2008}.
However, the typical analytical treatment, the Thomas-Fermi approximation (TFA), \cite{S_ho_binary_1996} which ignores the kinetic energy term, is not reliable in our case.
It is known that TFA cannot be used in the case of high density \cite{S_pu_properties_1998}, which is the case we are studying.
Instead, we numerically solve the GPE:
\begin{equation}
i\hbar\frac{\partial}{\partial t}\psi_{i} = \left[-\frac{\hbar^{2}}{2m}\nabla^{2}+V_{i}+\sum_{j=a,b}U_{ij}(N_{i}-\delta_{ij})|\psi_{j}|^{2}\right]\psi_{i}\label{eq: GPE}
\end{equation}
where $\delta_{ij}$ is the Kronecker delta which cannot be ignored if $N_i$ is of order one; $\psi_{i}$ and $N_{i}$ are the single mode wavefunction and the number of particles of the $i$-th BEC component respectively.
The normalization is $\int d^3r |\psi_i|^2=1$ and the density is given by $\rho(r)=N_i|\psi_i(r)|^2$. The trapping potential is $V_{i}=m\omega_{i}^{2}r^{2}/2$, with trapping strength $\omega_{i}$, and
the interaction strength is $U_{ij}=4\pi\hbar^{2}a_{ij}/m$, with scattering length $a_{ij}$ between component $i$ and $j$.
Our target is to find the ground state energy and wavefunction, which can be done by using the imaginary time method \cite{S_chiofalo_ground_2000}.
First, we use a Wick rotation $t \to -it$ on Eq. (\ref{eq: GPE}) to obtain the corresponding diffusion equation,
which is then reduced to two coupled 1D non-linear diffusion equations with the assumption of spherical symmetry.
Finally, we let the system relax to the ground state with the fourth order Runge-Kutta method in time and finite difference method in space.
After finding the ground state wavefunction, we can use it to calculate the mean-field energy functional:
\begin{equation}
E[\psi_{a},\psi_{b};N_{a},N_{b}] = \sum_{i=a,b}N_{i}\int d^{3}r\left(\frac{\hbar^{2}}{2m}|\nabla\psi_{i}|^{2}+V_{i}|\psi_{i}|^{2}+\frac{1}{2}(N_{i}-1)U_{ii}|\psi_{i}|^{4}\right)+N_{a}N_{b}\int d^{3}rU_{ab}|\psi_{a}|^{2}|\psi_{b}|^{2}\label{eq: energy-functional}
\end{equation}
which depends on the spatial mode $\psi_i$ and the number of particles $N_i$. Note that the spatial modes $\psi_i$ depend implicitly on $N_i$ through Eq. (\ref{eq: GPE}).
In our scheme, the focus is the ground state energy $E(N,n)$ as a function of $N_a=N-n$ and $N_b=n$ because the total number of particles $N=N_{a}+N_{b}$ in the two-component BEC is fixed.
After solving a set of BECs with different small component in the range $n\in[0,200]$,
we fit the results up to fourth order to get the expansion coefficients
\begin{equation}
\frac{1}{\hbar}E(N,n) = \eta(N,n)=\eta_{0}(N)+\eta_{1}(N)n+\eta_{2}(N)n^{2}+\eta_{3}(N)n^{3}+\eta_{4}(N)n^{4}+... \label{eq: etas}
\end{equation}

Fig. \ref{cap: Properties} shows how the most relevant properties of the ground state of the two-component BEC change with $\omega_b$.
For the scheme described in the main text, the interesting regime is when component B is located at the center of the trap.
This can be achieved with a slightly higher trapping for $\omega_b$ in this weakly phase separated regime as described in Fig. \ref{cap: Properties} with cat size $\bar{n}=100$.
Note that in the case of equal trapping $\omega_a=\omega_b$, the small component B will locate outside of component A because of the effective repulsion in this regime.
As shown in Fig. \ref{cap: Properties}a, the width of $\rho_b$ is close to the width of a Gaussian at around $\omega_b/2\pi=50 \mbox{Hz}$, while at higher $\omega_b$, the width is larger than the corresponding Gaussian because of the self-repulsion with other atoms in the same component B.
The same effects can be observed for the real density $\rho_b(r=0)$ at the center (Fig. \ref{cap: Properties}b), which is lower than the corresponding Gaussian density with $\omega_b$.
When $\omega_b/2\pi\gg250 \mbox{Hz}$, the component B has higher density than the main component A. This allows us to ignore most effects of the component A, including the quantum depletion.
Fig. \ref{cap: Properties}c-e shows the expansion coefficients $\eta_k$.
Note that both $\eta_2$ and $\eta_3$ have zero-crossing points.
With zero Kerr coefficient, $\eta_2=0$, the system may be used to store spin states for a long time.
Also, the zero third order, $\eta_3=0$, suggests a regime to create good small spin cat states.
Fig. \ref{cap: Properties}f shows the effects of the third order term on the shift of the ``best'' cat time $\tau_c^*$, see definition below.

Qualitatively, the change in $\eta_2$ with respect to $\omega_b$ can be understood as follow. The contributions to the Kerr nonlinearity come from intra-species (aa, bb) and inter-species (ab) interactions, which have opposite sign to each other. When the trapping is weak and identical for both components, the Kerr nonlinearity is close to zero. Also, for the phase separated regime, the component B is staying in the outer region. When the trapping frequency $\omega_b$ for the B component is increased, the B component moves to the center and the overlap between A and B increases at first, which leads to an increase in the inter-species interaction term, resulting in a larger and negative Kerr nonlinearity. For very strong trapping of the B component, the overlap between A and B decreases again whereas the intra-species interaction for the B component increases strongly, leading to a large positive Kerr nonlinearity. This explains the crossover from negative to positive Kerr nonlinearity as shown in Fig. S1d. In contrast, for the non-phase separated regime, the B component always stays inside the A component, and there is no crossover as discussed in Section IV (see Fig. S5).

\section{Ground state energy from first order perturbation theory}

\begin{figure}
\begin{centering}
\includegraphics[width=0.7\textwidth]{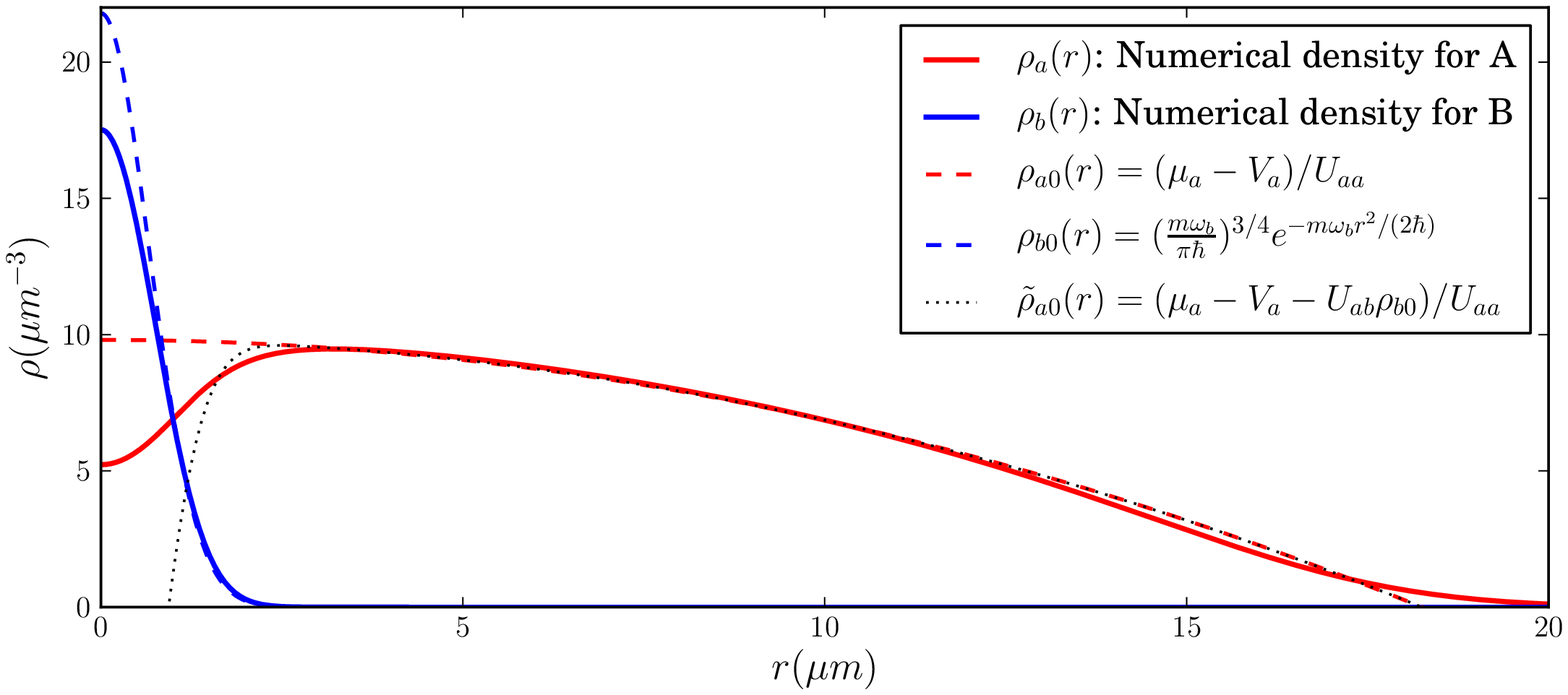}
\par\end{centering}
\caption{\label{cap: wavefunc-demo}
Numerical density distribution for both components (A and B) and its approximation with cat size $\bar{n}=100$ and trapping strength $\omega_b/2\pi = 500 \mbox{Hz}$. $\rho_{a0}(r)$ and $\rho_{b0}(r)$ are the unperturbed wavefunction used by the first order perturbation calculation. $\tilde{\rho_{a0}}(r)$ is another approximation. See text for details.
Other parameters used are the same as in Fig. \ref{cap: Properties}.
}
\end{figure}

The numerically obtained spatial density distribution $\rho_i$ is shown in Fig. \ref{cap: wavefunc-demo}.
The approximate solution of a harmonic oscillator ground state  $\rho_{b0}$ for B is good.
If we follow a Thomas-Fermi approach similar to the one used in the previous paper \cite{S_Rispe} by dropping the kinetic energy term in Eq. (\ref{eq: GPE}), we will get $\tilde{\rho}_{a0}=(\mu_{a}-V_{a}-U_{ab}\rho_{b0})/U_{aa}$.
As expected, this approximation is not good and gives a negative density as shown in Fig. \ref{cap: wavefunc-demo}.
In contrast, the TFA solution for a single component BEC $\rho_{a0}$ gives a fair approximation for A, given by \cite{S_pitaevskii_bose-einstein_2003,S_pethick_bose-einstein_2008}:
\begin{equation}
\begin{aligned}
\phi_{a0}(\mathbf{r};N_{a}) & = \sqrt{\frac{\mu_{a0}(N_{a})-V_{a}}{N_{a}U_{aa}}},\\
\phi_{b0}(\mathbf{r}) & = \left(\frac{m\omega_{b}}{\pi\hbar}\right)^{3/4}e^{-m\omega_{b}r^{2}/(2\hbar)},\\
\end{aligned}
\begin{aligned}
\mu_{a0}(N_{a}) & = \frac{1}{2}\hbar\omega_{a}\left(\frac{15a_{a}}{\sqrt{\hbar/(m\omega_{a})}}\right)^{2/5}N_{a}^{2/5}\\
\mu_{b0} & = \frac{3}{2}\hbar\omega_{b}
\end{aligned}
\label{eq: GPE-unpert-sol}
\end{equation}
Therefore we perform first order perturbation theory with the following splitting for the GPE:
\begin{eqnarray}
i\hbar\frac{\partial}{\partial t}\psi_{a} & = & (\underset{H_{a0}}{\underbrace{-\frac{\hbar^{2}}{2m}\nabla^{2}+V_{a}+N_{a}U_{aa}|\phi_{a}|^{2}}}+\underset{H_{a1}}{\underbrace{N_{b}U_{ab}|\phi_{b}|^{2}}})\psi_{a}\\
i\hbar\frac{\partial}{\partial t}\psi_{b} & = & (\underset{\mathcal{H}_{b0}}{\underbrace{-\frac{\hbar^{2}}{2m}\nabla^{2}+V_{b}}}+\underset{\mathcal{H}_{b1}}{\underbrace{N_{a}U_{ab}|\phi_{a}|^{2}+(N_{b}-1)U_{bb}|\phi_{b}|^{2}}})\psi_{b}
\end{eqnarray}
where $\mathcal{H}_{i0}$ is the unperturbed Hamiltonian and the perturbation
is given by $\mathcal{H}_{i1}$. Note that $N_{a}-1 \approx N_{a}$ is used.
The solutions of $\mathcal{H}_{i0}$ are given by Eq. (\ref{eq: GPE-unpert-sol}).

To calculate the energy analytically, we expand the ground state energy $E(N_{a},N_{b})$ as the Taylor series:
\begin{eqnarray}
E(N_{a},N_{b}) & = & E(\bar{N}_{a},\bar{N}_{b})+\sum_{i=a,b}\left.\frac{\partial E}{\partial N_{i}}\right|_{(\bar{N}_{a},\bar{N}_{b})}(N_{i}-\bar{N}_{i})+\frac{1}{2}\sum_{j=a,b}\sum_{i=a,b}\left.\frac{\partial}{\partial N_{j}}\frac{\partial E}{\partial N_{i}}\right|_{(\bar{N}_{a},\bar{N}_{b})}(N_{i}-\bar{N}_{i})(N_{j}-\bar{N}_{j})+...
\end{eqnarray}
Note that the chemical potentials (energy change with respect to the number of particles) are given by $\mu_{i}(N_{a},N_{b})=\frac{\partial E}{\partial N_{i}}(N_{a},N_{b})$.
Since the main component $A$ in the scheme is much larger than the small component $B$, or $N-n\gg n$, the expansion can be carried out around the point $(N,0)$ :
\begin{eqnarray}
\hbar\eta_{0}(N) & = & E(N,0)\\
\hbar\eta_{1}(N) & = & -\mu_{a}(N,0)+\mu_{b}(N,0)\label{eq: eta1-def}\\
\hbar\eta_{2}(N) & = & \frac{1}{2}\left[\partial_{N_{a}}\mu_{a}(N,0)-\partial_{N_{b}}\mu_{a}(N,0)-\partial_{N_{a}}\mu_{b}(N,0)+\partial_{N_{b}}\mu_{b}(N,0)\right]\label{eq: eta2-def}
\end{eqnarray}
Note that the $\mu_i$ here denote the exact chemical potentials from the GPE, which can be approximated by the unperturbed $\mu_{i0}$ plus the perturbed chemical potential $\Delta\mu_i$:
\begin{equation}
\mu_{i} = \mu_{i0}+\Delta\mu_{i}
\end{equation}
Using the unperturbed solutions Eq. (\ref{eq: GPE-unpert-sol}), the chemical potential can be calculated as:
\begin{eqnarray}
\Delta\mu_{a} & = & U_{ab}N_{b}\left\langle \phi_{a0}\left||\phi_{b0}|^{2}\right|\phi_{a0}\right\rangle \\
 & = & \frac{N_{b}}{N_{a}}\left(\frac{U_{ab}}{U_{aa}}\mu_{a0}(N_{a})-\frac{3}{4}\frac{U_{ab}\omega_{a}}{U_{aa}\omega_{b}}\hbar\omega_{a}\right)\\
\Delta\mu_{b} & = & U_{ab}N_{a}\left\langle \phi_{b0}\left||\phi_{a0}|^{2}\right|\phi_{b0}\right\rangle +U_{bb}(N_{b}-1)\left\langle \phi_{b0}\left||\phi_{b0}|^{2}\right|\phi_{b0}\right\rangle \\
 & = & \left(\frac{U_{ab}}{U_{aa}}\mu_{a0}(N_{a})-\frac{3}{4}\frac{U_{ab}\omega_{a}}{U_{aa}\omega_{b}}\hbar\omega_{a}\right)+U_{bb}(N_{b}-1)(\sqrt{2}s_{b})^{-3}
\end{eqnarray}
where $s_{i}=\sqrt{\pi\hbar/(m\omega_{i})}$ is the characteristic length of a Gaussian.
Note that the perturbation involves an integration whose range is chosen to be the whole space for simplicity, which is justified by the fact that component B is much narrower than component A when $\omega_{b}\gg\omega_{a}$ (see Fig. \ref{cap: wavefunc-demo}).
Substituting these results back into $\eta_{1}$ in Eq. (\ref{eq: eta1-def}), we have:
\begin{equation}
\hbar\eta_{1}(N)=-\mu_{a0}(N)+\underset{\mu_{b0}}{\underbrace{\frac{3}{2}\hbar\omega_{b}}}+\underset{\Delta\mu_{b}(N,0)}{\underbrace{\frac{U_{ab}}{U_{aa}}\mu_{a0}(N)-\frac{3}{4}\frac{U_{ab}\omega_{a}}{U_{aa}\omega_{b}}\hbar\omega_{a}-U_{bb}(\sqrt{2}s_{b})^{-3}}}\label{eq: eta1-approx}
\end{equation}
The third and fourth terms on the right hand side are the effective interaction
between the main BEC and the component $B$. The last term is the repulsion
between the particles in component $B$. The fourth term is small
when $\omega_{b}\gg\omega_{a}$ and can be ignored. This result
gives a very good approximation as demonstrated in Fig. \ref{cap: Properties}c.

Similarly, differentiating the chemical potential yields the second order term $\eta_{2}$ in Eq. (\ref{eq: eta2-def}):
\begin{eqnarray}
\hbar\eta_{2}(N) & = & \frac{1}{2}\left[\underset{\partial_{N_{a}}\mu_{a}(N,0)}{\underbrace{\frac{2}{5}\frac{\mu_{a0}(N)}{N}}}-\underset{\partial_{N_{b}}\mu_{a}(N,0)}{\underbrace{\left(\frac{U_{ab}}{U_{aa}}\mu_{a0}(N)-\frac{3}{4}\frac{U_{ab}\omega_{a}}{U_{aa}\omega_{b}}\hbar\omega_{a}\right)\frac{1}{N}}}-\underset{\partial_{N_{a}}\mu_{b}(N,0)}{\underbrace{\frac{2}{5}\frac{U_{ab}}{U_{aa}}\frac{\mu_{a0}(N)}{N}}}+\underset{\partial_{N_{b}}\mu_{b}(N,0)}{\underbrace{U_{bb}(\sqrt{2}s_{b})^{-3}}}\right] \label{eq: eta2-approx}
\end{eqnarray}
The first three derivatives are smaller than the last term when $\omega_{b}\gg\omega_{a}$ and $N\to\infty$.
Therefore, at high $\omega_{b}$, the last term dominates $\eta_{2}(N)$, yielding 
\begin{equation}
\hbar\eta_{2}(N)\approx \frac{U_{bb}}{2}(\sqrt{2}s_{b})^{-3}=\frac{U_{bb}}{2}\left(\frac{m\omega_{b}}{2\pi\hbar}\right)^{3/2}.\label{eq: eta2-approx-simple}
\end{equation}
As shown in Fig. \ref{cap: Properties}b, Eq. (\ref{eq: eta2-approx-simple}) gives an order of magnitude estimation of $\eta_{2}(N)$.
Note that we can also get the dominant term as calculated above by assuming component A to have a constant density distribution $|\psi(r)|^2=\mu_{a0}/(N_aU_{aa})$, at $\omega_b\gg\omega_a$.
A better approximation should take into account the change in density $\rho_a$ as shown in Fig. \ref{cap: wavefunc-demo}.

\section{Effects of higher-order nonlinearities}

\begin{figure}
\begin{centering}
\includegraphics[width=0.7\textwidth]{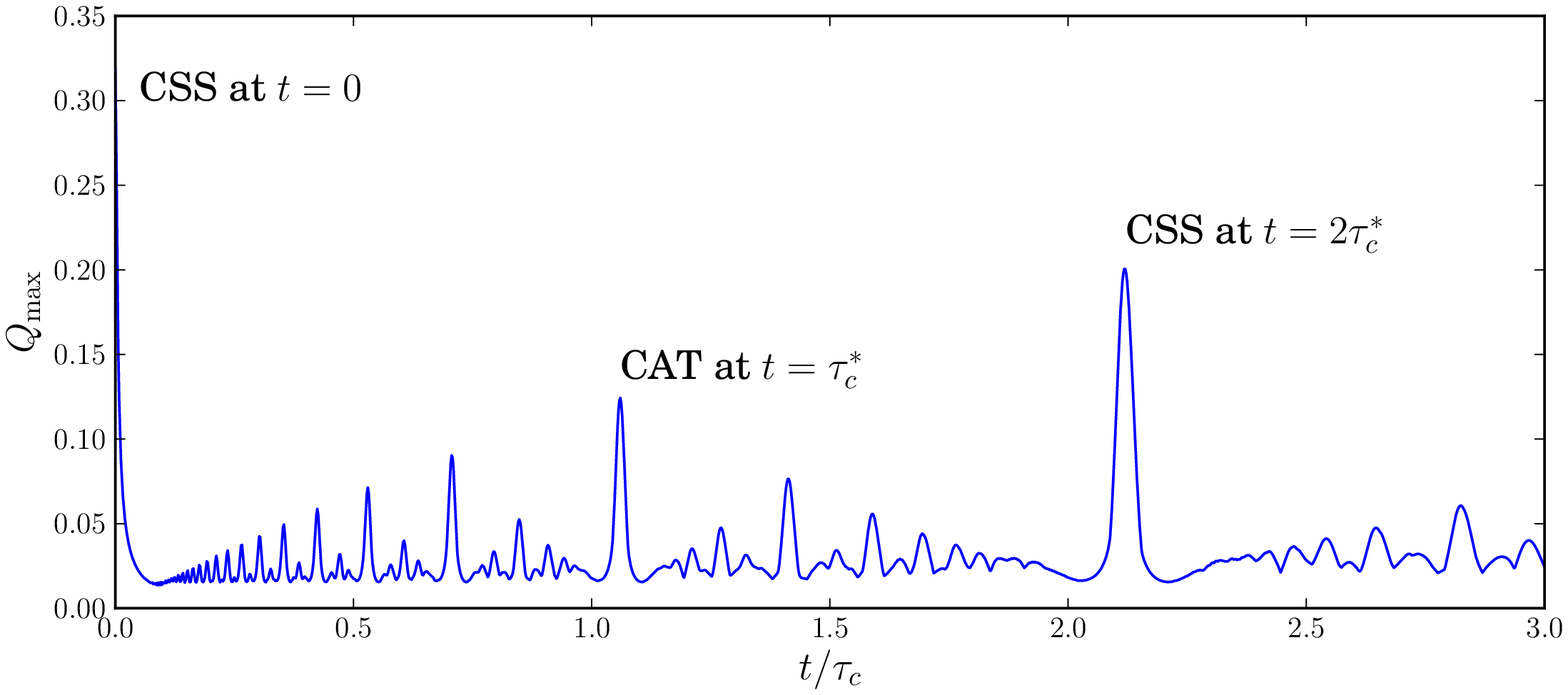}
\par\end{centering}
\caption{\label{cap: Qmax-tauc}
The maximum of the Q function, $Q_{\max}(\beta)$, as a function of the relative time $t/\tau_c$.
The ``best'' real cat time $\tau_{c}^{*}$ is defined as the time in which there are the two highest peaks in the $Q$-function.
The leftmost peak corresponds to the initial coherent spin state (CSS), with a value $Q_{\max}=1/\pi$.
The peak at $\tau_c^*/\tau_c \approx 1.06$ corresponds to the spin cat state (CAT). The time at which the CAT state is observed is shifted with respect to the ideal case $\tau_c^*/\tau_c=1$ because of the higher-order nonlinearities.
The highest peak at $\tau_{c}^{*}/\tau_{c}\approx2.12$ corresponds to the CSS at the revival time.
Note that all fitting orders are included for determining $\tau_{c}^{*}$.
The CAT state at $\tau_c^*$ and CSS at $2\tau_c^*$ are plotted in Figs. 3a and 3b in the main text.
The other parameters used are the same as in Fig. \ref{cap: Properties}.
}
\end{figure}

\begin{figure}
    \centering
    \subfigure[$\bar{n}=9$ and $\tau_{c}^{*}=1.005 \tau_{c}$]{\includegraphics[width=0.32\textwidth]{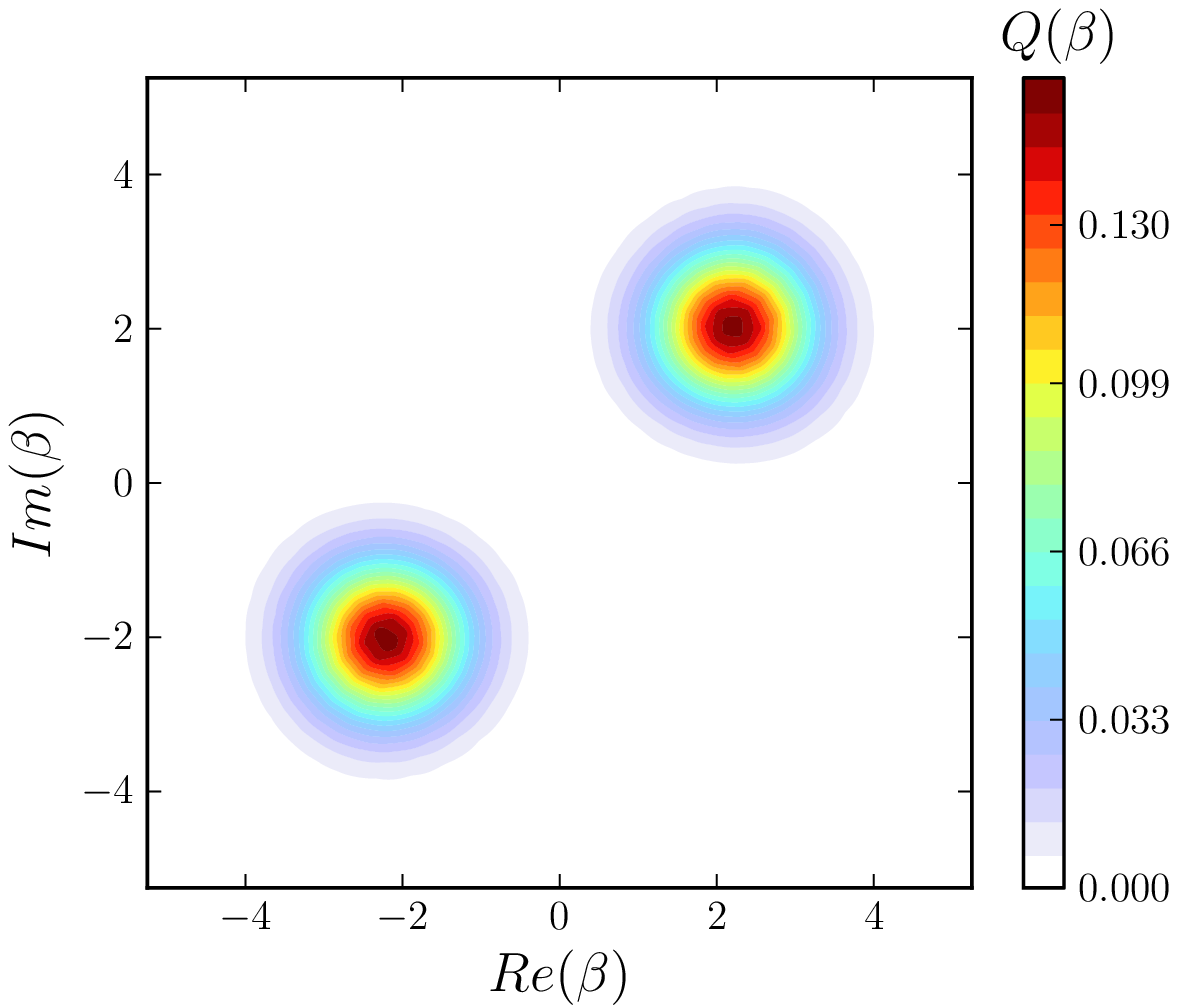}}
    \subfigure[$\bar{n}=49$ and $\tau_{c}^{*}=1.026 \tau_{c}$]{\includegraphics[width=0.32\textwidth]{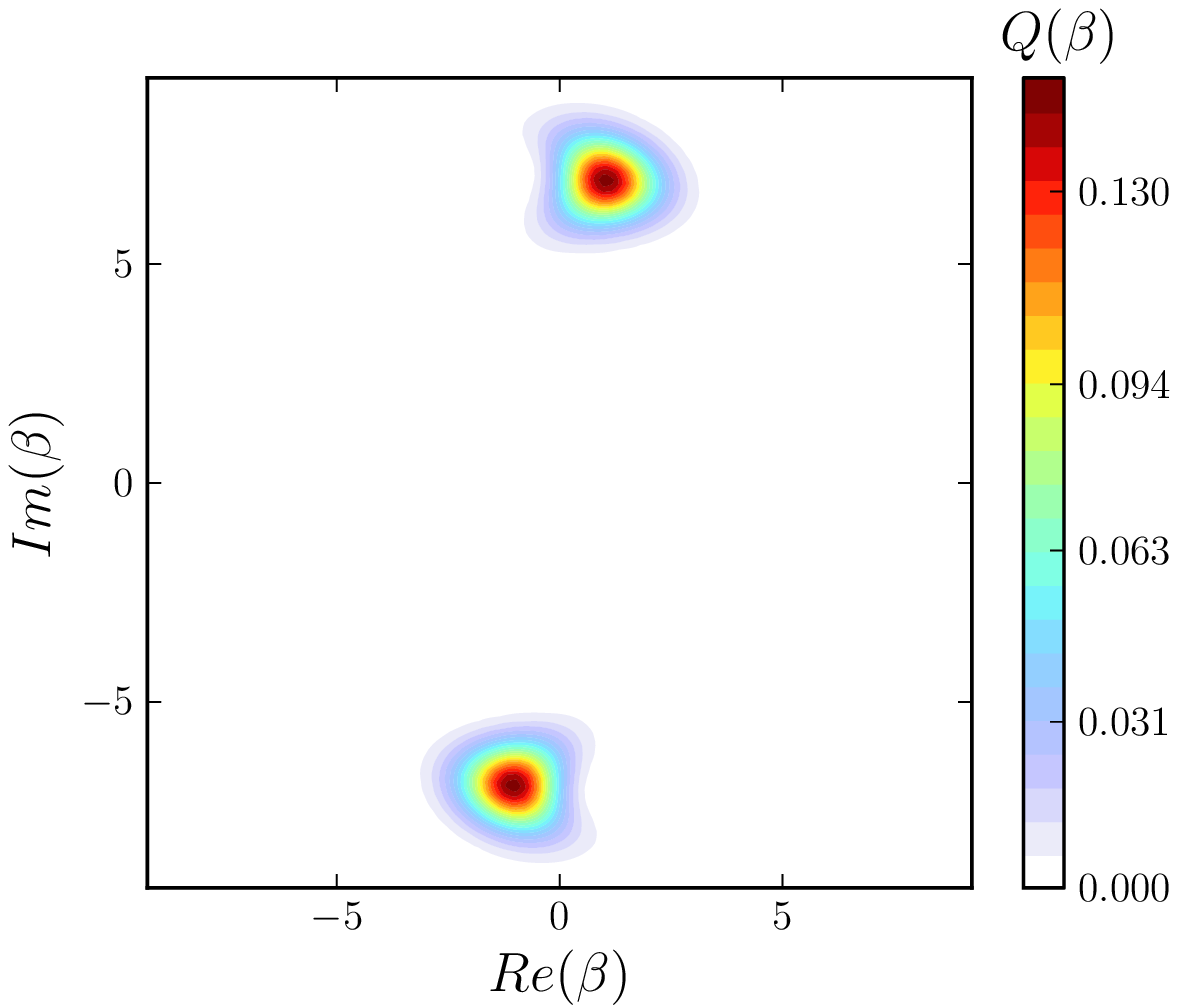}}
    \subfigure[$\bar{n}=400$ and $\tau_{c}^{*} = 1.52 \tau_{c}$]{\includegraphics[width=0.32\textwidth]{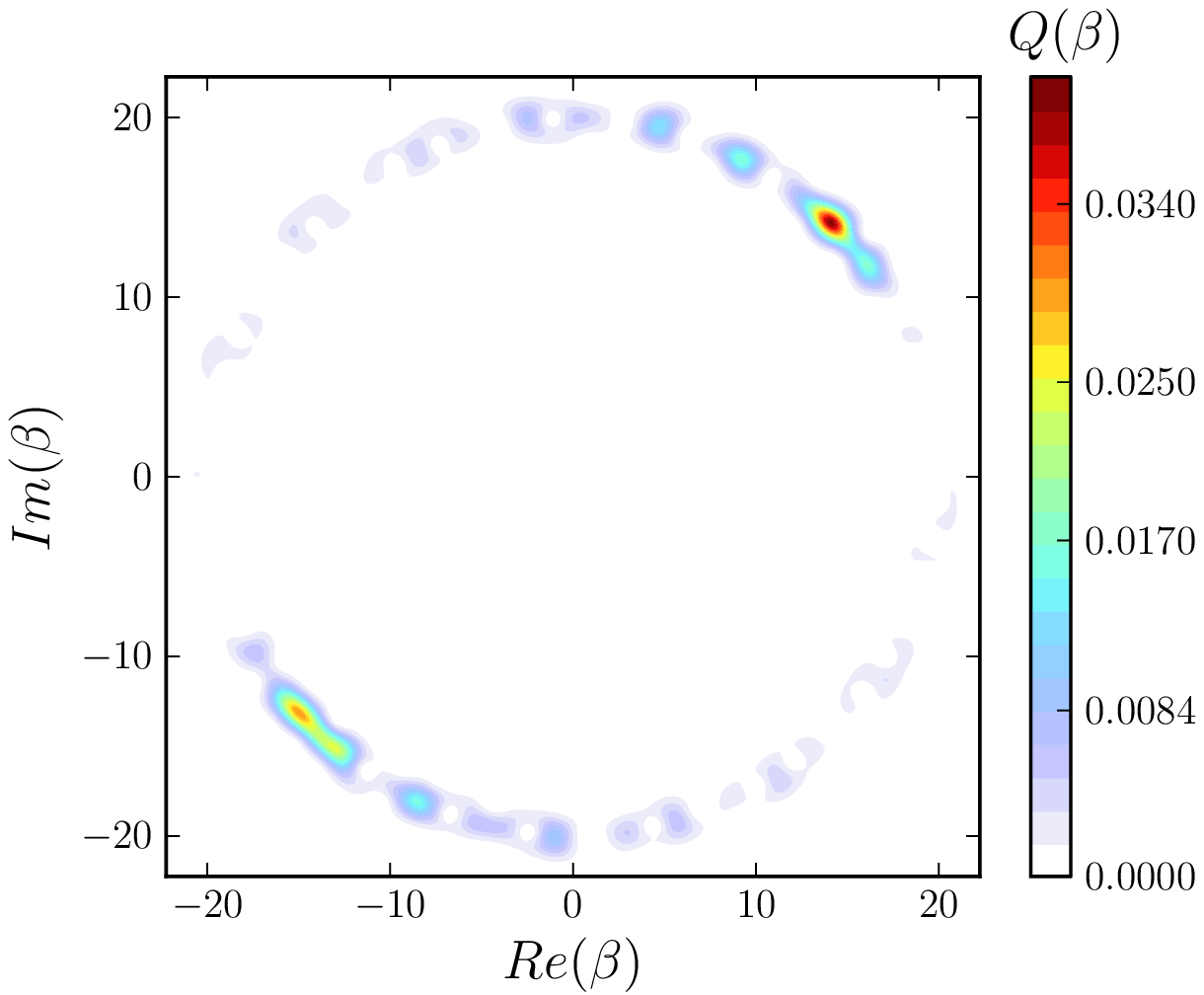}}
\caption{\label{cap: Q-function-nbar}
Plot of $Q(\beta,\tau_c^*)$ for different cat sizes $\bar{n}$.
(a) $\bar{n}=9$. The third order nonlinearity $\eta_3$ is weak, so the Q function looks like a perfect circle.
(b) $\bar{n}=49$. The effects of $\eta_3$ begin to appear and the cat state is distorted slightly.
(c) $\bar{n}=400$. Both $\eta_3$ and $\eta_4$ are significant. The two peaks are distorted and not symmetric.
Note that $\tau_{c}^{*}$ is not quite well defined in this case.
The case $\bar{n}=100$ is plotted in Fig. 3a in the main text with $\omega_{b}=2\pi 500 \mbox{Hz}$ and $\tau_{c}=0.646 \mbox{s}$.
The other parameters used are the same as in Fig. \ref{cap: Properties}.
}
\end{figure}

Cat states can be distorted by higher order nonlinearities. Thus we need to find out to what extent the cat states are distorted and whether the distortion is tolerable.
Another practical problem is to figure out the optimal time to observe a cat state in real experiments. We define the ``best'' cat time $\tau_c^*$ as the time with the two highest peaks in Q function. This definition is based on the feature of the cat states that two separated peaks in the Q function should be distinguished clearly.

This method is illustrated by Fig. \ref{cap: Qmax-tauc} with the highest peak value $Q_{max}$ plotted over time.
It is clear that the peak for the cat state is located near $\tau_c^*/\tau_c=1$ as expected.
In practice, we search around the nearby region, say $\tau_c^*/\tau_c \in [0.8,1.6]$, for the highest peak. The resulting $\tau_c^*$ corresponds to the best cat time.
We further manually check that there are indeed only two opposite peaks in phase space.
The resulting shift in the cat time is plotted in Fig. \ref{cap: Properties}d.
Note that the $\tau_c^*$ depends on $\bar{n}$, see also Fig. \ref{cap: Properties}f.

In the scheme, the output light is of the form $|\chi(t)\rangle_{L}=\sum_{n}c_{n}e^{-i\eta(N,n)t}|n\rangle_L$ with the initial condition $|\chi(0)\rangle_{L}=|\alpha\rangle_{L}$ and $\alpha=\sqrt{\bar{n}}$.
Hence, the Q-function without loss is
\begin{equation}
Q(s,\theta,t)=\frac{1}{\pi}e^{-(\alpha-s)^{2}}\left|\sum_{n}\left(\frac{(\alpha s)^{n}}{n!}e^{-\alpha s}\right)e^{-in\theta}e^{-i\eta(N,n)t}\right|^{2}\label{eq: Q-function-eta}
\end{equation}
where the phase space is defined by $\beta=se^{i\theta}$.
This equation is numerically evaluated to obtain the Q-function for given $\eta_k$, which are obtained by fitting the solutions of the GPE Eq. (\ref{eq: GPE}).
A few more figures corresponding to Fig. 3a in main text are plotted in Fig. \ref{cap: Q-function-nbar} for different cat sizes $\bar{n}$.
 One can see that the higher order effects ($k\ge3$) are weak for small $\bar{n}$, but significant for larger $\bar{n}$.

\section{Phase separated regime and non-phase separated regime}

\begin{figure}
    \centering
    \subfigure[$\eta_2(\omega_b)$ for different $a_{ab}$]{\includegraphics[width=0.28\textwidth]{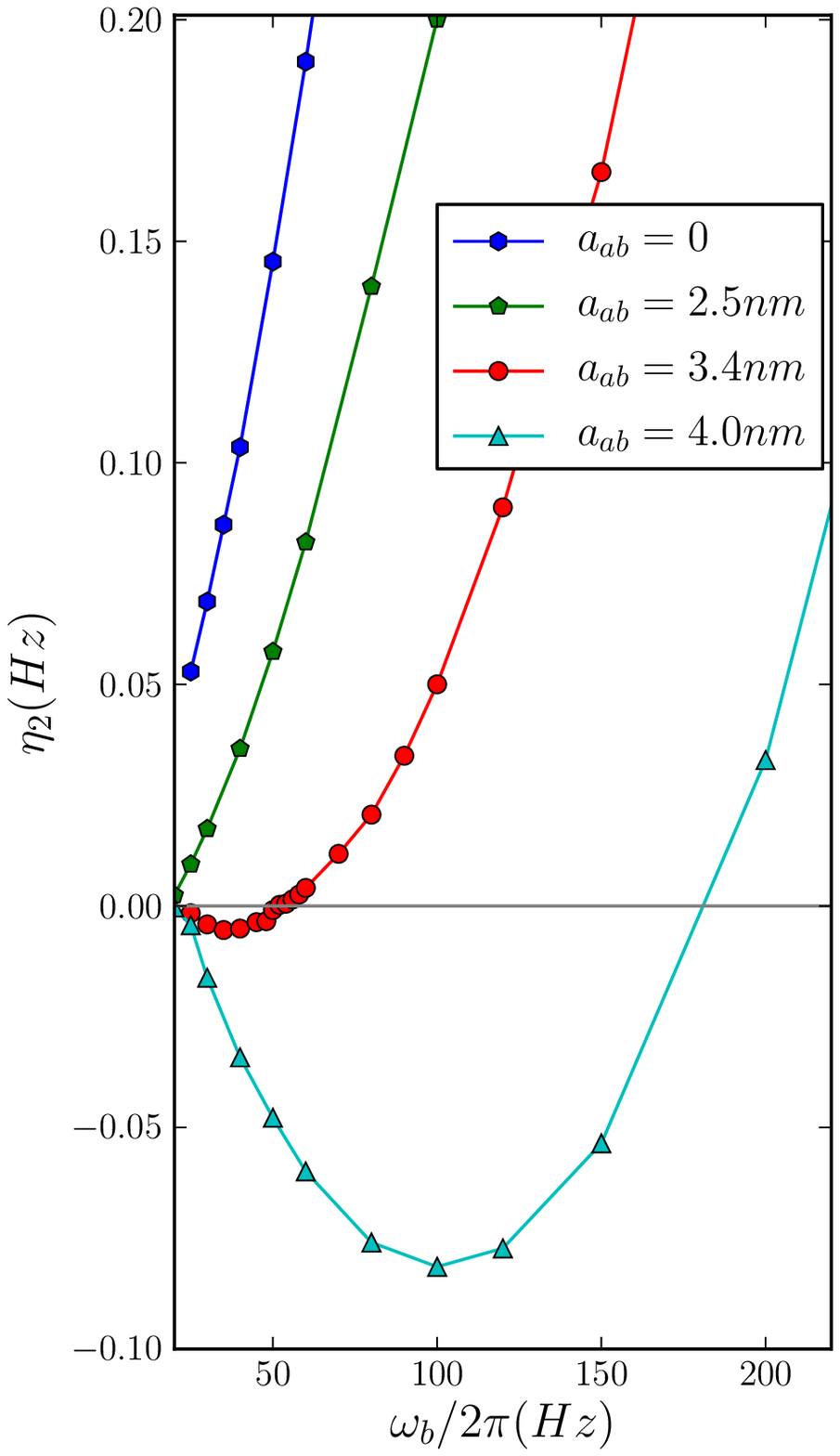}}
    \subfigure[$\eta_3(\omega_b)$ for different $a_{ab}$]{\includegraphics[width=0.28\textwidth]{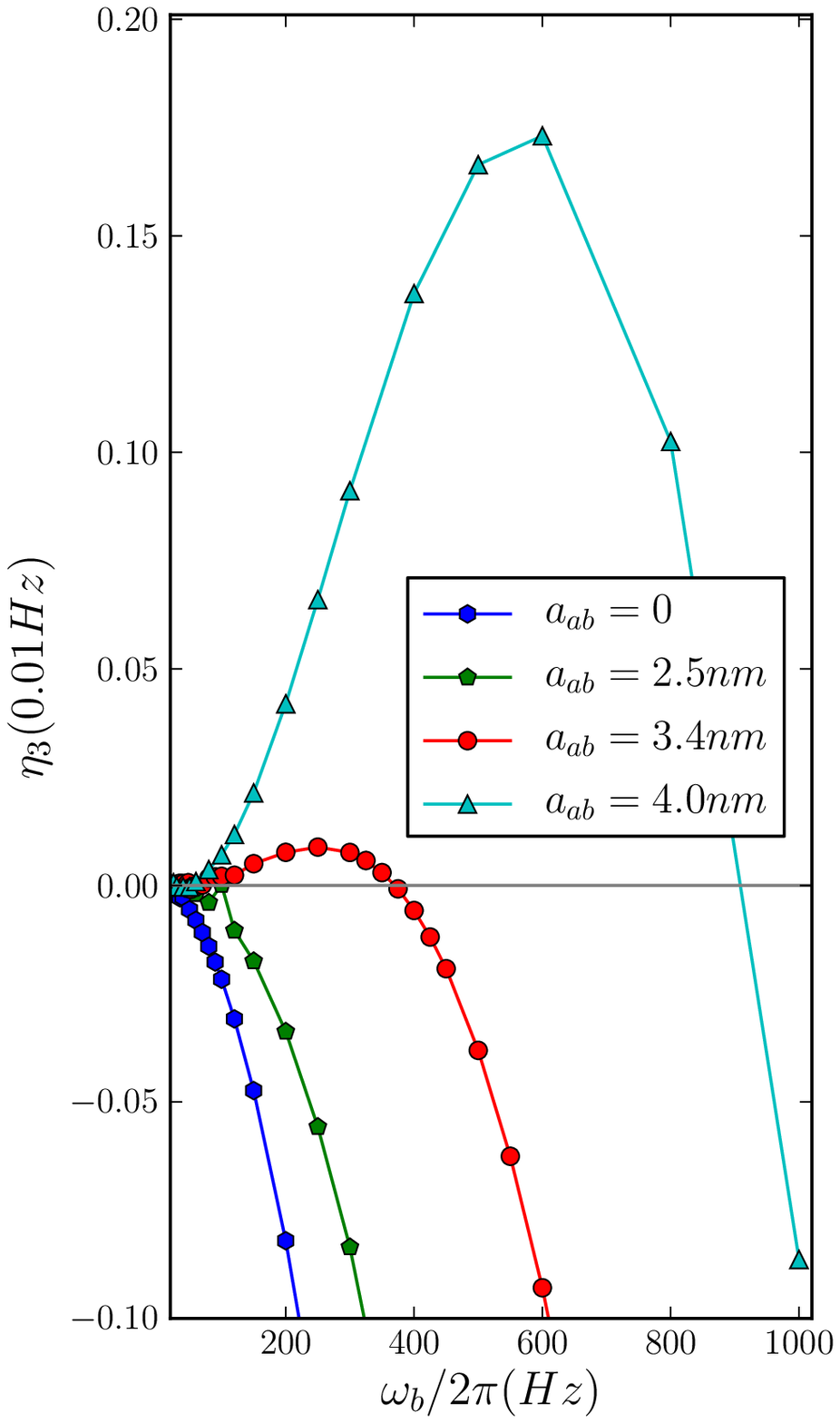}}
    \subfigure[$\eta_4(\omega_b)$ for different $a_{ab}$]{\includegraphics[width=0.28\textwidth]{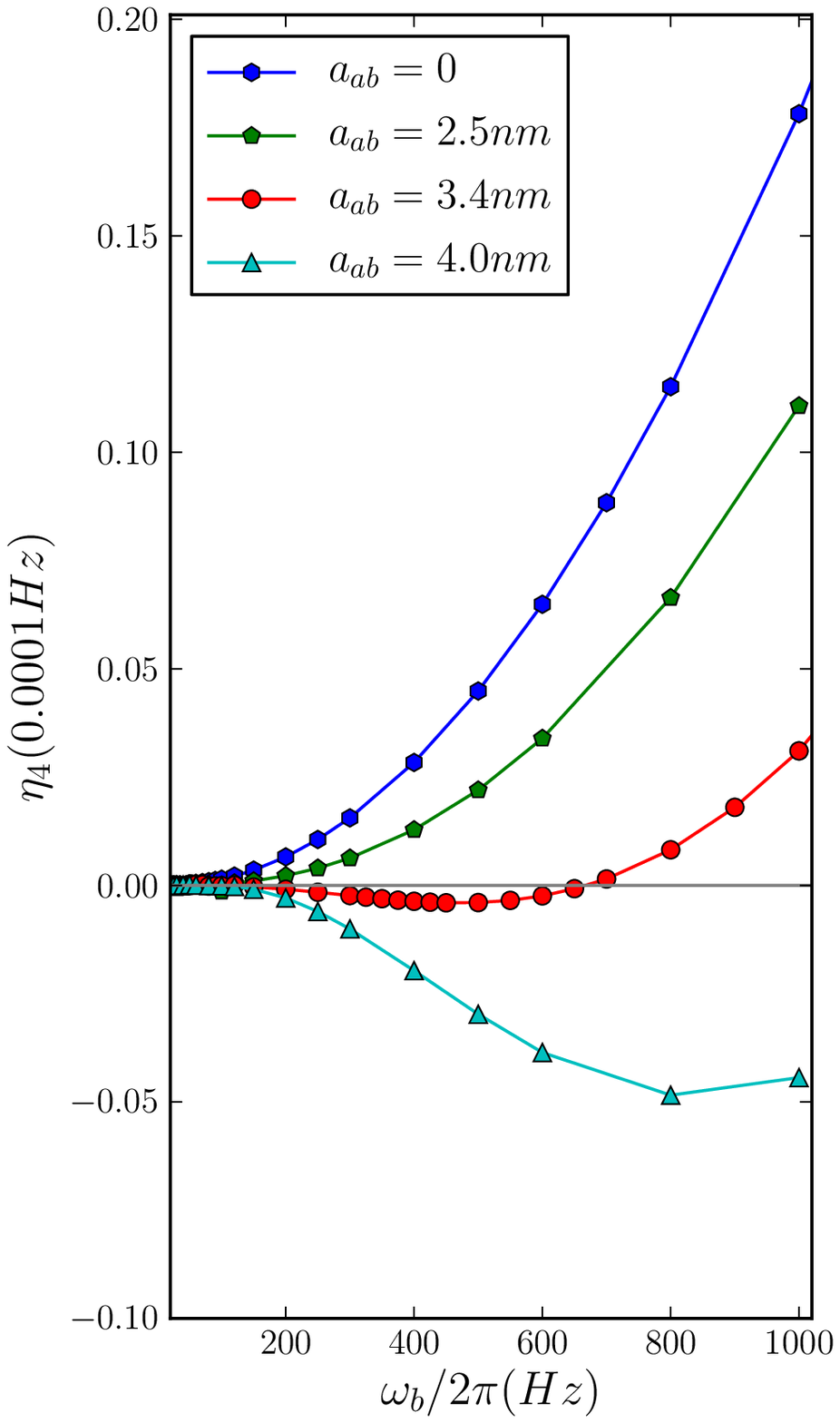}}
\caption{\label{cap: Na-varying-aab}
Effects of the cross-scattering length $a_{ab}$ on (a) $\eta_{2}$, (b) $\eta_{3}$, (c) $\eta_{4}$, with $a_{aa}=2.8 \mbox{nm}$ and $a_{bb}=3.4 \mbox{nm}$
 ($a_{ii}$ is the self-scattering length of component $i$).
Both $a_{ab}=0$ and $a_{ab}=2.5 \mbox{nm}$ are in the non-phase separated regime $a_{ab}^{2}<a_{aa}a_{bb}$,
while $a_{ab}=3.4 \mbox{nm}$ and $a_{ab}=4.0 \mbox{nm}$ are in the phase separated regime $a_{ab}^{2}>a_{aa}a_{bb}$.
When $a_{ab}$ is turned on gradually, the magnitude of all nonlinear coefficients $\eta_{k}$ decreases at first because the effective scattering length for the two components decreases.
All coefficients show a qualitative change, with a zero-crossing point in the phase separated regime.
Compared with the non-phase separated regime, say, $a_{ab}=0$, the phase separated regime can have a relatively weak higher-order effect even for high trapping frequencies, e.g. the small $\eta_3$ at $\omega_{b}/2\pi= 500 \mbox{Hz}$ which is used in Fig. 3 of the main text.
Note that the y axis is rescaled by factors of 100 from left to right for easy comparison.
The parameters used are the same as in Fig. \ref{cap: Properties}, except $a_{ab}$.
}
\end{figure}

The scheme should also work in the non-phase separated regime $a_{ab}^{2} < a_{aa}a_{bb}$. Fig. \ref{cap: Na-varying-aab} shows the coefficients $\eta_2$, $\eta_3$, $\eta_4$ for different values of the inter-species scattering length $a_{ab}$, with $a_{aa}=2.8 \mbox{nm}$ and $a_{bb}=3.4 \mbox{nm}$.
The plots suggest that the Kerr effect is also strong in the non-phase separated regime, but the higher order terms might limit the resulting cat size $\bar{n}$.
The main qualitative difference is that there are no zero-crossing points for $\eta_k$ in the non-phase separated regime.
These results further suggest that the weakly phase separated regime is advantageous because the higher-order terms are very small around $\omega_b/2\pi \approx 400 \mbox{Hz}$.

\section{Atom loss rates}

\begin{figure}
\begin{centering}
\includegraphics[width=0.57\textwidth]{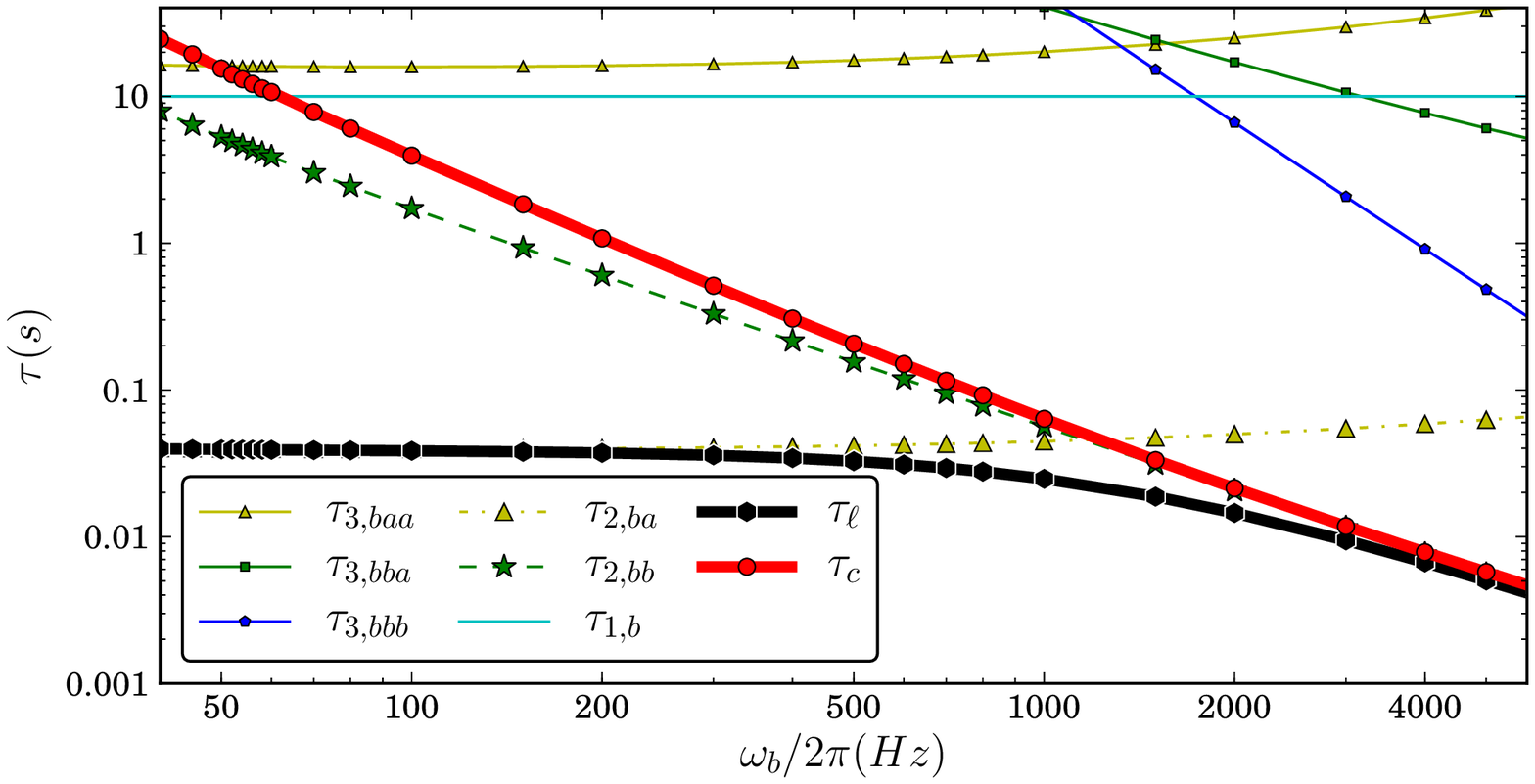}
\par\end{centering}
\caption{\label{cap: Rb-loss}
Cat time $\tau_{c}$ and one-atom loss time $\tau_{\ell}$ for Rubidium with $\bar{n}=10$, $N=10^{5}$ and non-zero two-body loss rate $L_{2,ij}$.
It is clear that the time to lose one atom through the two-body loss within the same component, $\tau_{2,bb}$, dominates at high $\omega_{b}$,
which has a similar scaling as the cat time $\tau_{c}\sim \omega_b^{-3/2}$. This limits
the maximum $\bar{n}$ to around 10 atoms. Parameters: Rubidium atoms
$^{87}$Rb with scattering length $a_{aa}=100.44r_{B}$, $a_{bb}=95.47r_{B}$,
$a_{ab}=88.28r_{B}$, where $r_{B}$ is the Bohr radius. The atom
loss rates are $L_{1}=0.01$/s, $L_{2,aa}=0$, $L_{2,bb}=119\times10^{-21}m^{3}$/s,
$L_{2,ab}=78\times10^{-21}$m$^{3}$/s, $L_{3}=6\times10^{-42}$m$^{6}$/s
\cite{S_li_spin_2009}, and $\omega_{a}/2\pi=20$Hz.
}
\end{figure}

\begin{figure}
\begin{centering}
\includegraphics[width=0.49\textwidth]{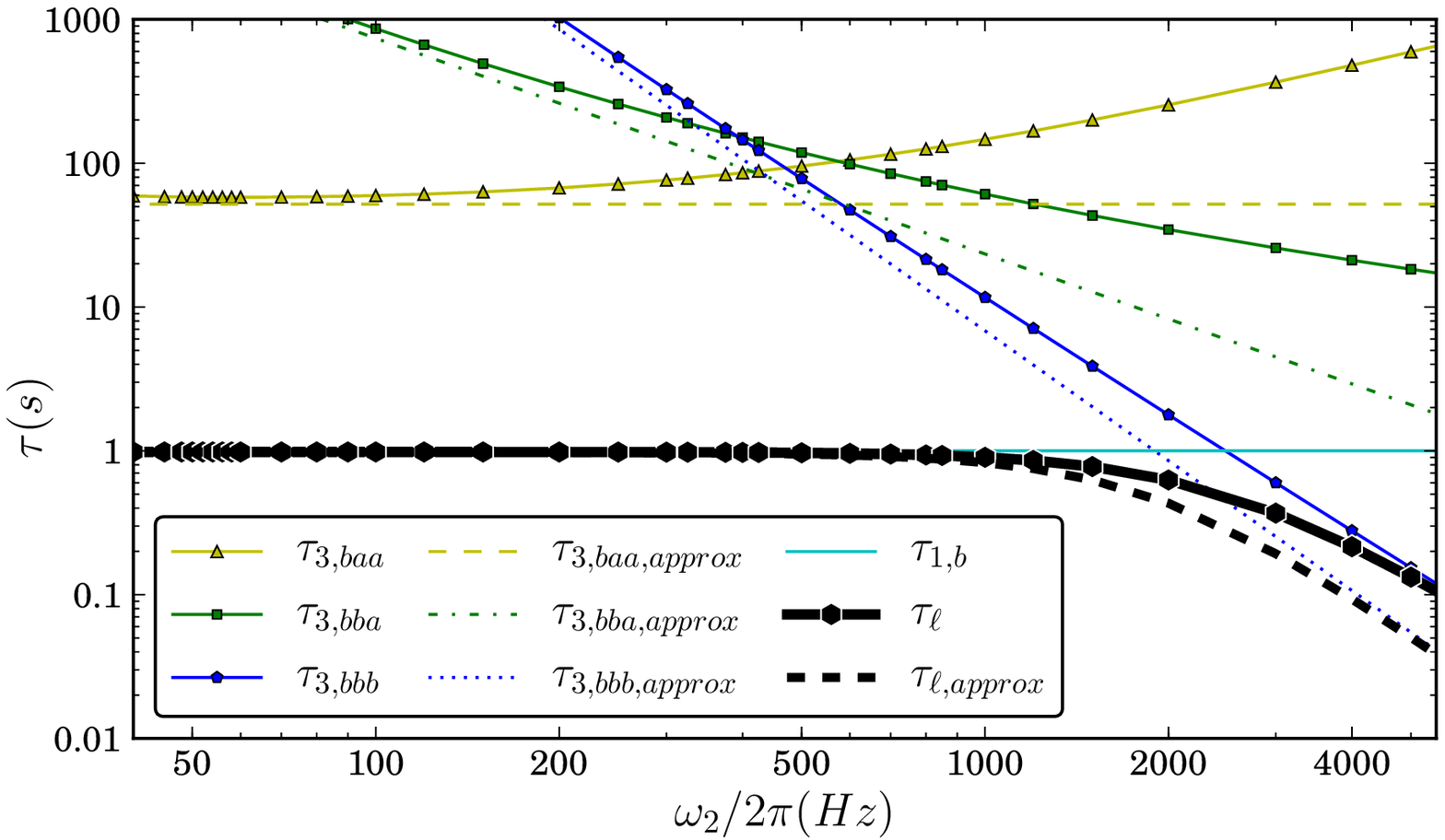}
\par\end{centering}
\caption{\label{cap: loss-approx}
Comparison of the numerical and approximation results for all loss processes corresponding to Fig. 2a in the main text.
The solid curves show the numerical solutions of the time to lose one atom $\tau_{m,c}$ through an $m$-body process with particle combination $c$, while the approximations are shown as dotted or dashed curves with the same color. The time to lose one atom through all processes is calculated by $\tau_\ell = (\tau_{1,b}^{-1} + \tau_{3,baa}^{-1} + \tau_{3,bba}^{-1} + \tau_{3,bbb}^{-1})^{-1}$.
Note that the approximations here are essentially lower bounds for the numerical solutions, as shown in this figure. See text for details.
}
\end{figure}

The atom loss rate for component $i$ is given by \cite{S_chin_feshbach_2010,S_li_optimum_2008,S_li_spin_2009}:
\begin{equation}
\frac{dN_{i}}{dt}=-\tau_{i,\ell}^{-1}=-\left(L_{1,i}\int d^{3}r\rho_{i}+\sum_{j=a,b}L_{2,ij}\int d^{3}r\rho_{i}\rho_{j}+\sum_{j=a,b}\sum_{k=a,b}L_{3,ijk}\int d^{3}r\rho_{i}\rho_{j}\rho_{k}\right) \label{eq: loss-rate-equation}
\end{equation}
where $L_{1,i}$, $L_{2,ij}$, $L_{3,ijk}$ are the one, two and three particle collision loss rates.
Note that the density $\rho_{i}=N_{i}|\psi_{i}|^{2}$ in the equation also depends on the numbers of particles $N_{i}$ which decrease over time.
As discussed in the main text, the individual times to lose one particle through an $m$-body process with particle combination $c$ are defined as $\tau_{1,i}=(L_{1,i}\int d^{3}r\rho_{i})^{-1}$, $\tau_{2,ij}=(L_{2,ij}\int d^{3}r\rho_{i}\rho_{j}))^{-1}$ and $\tau_{3,ijk}=(L_{3,ijk}\int d^{3}r\rho_{i}\rho_{j}\rho_{k})^{-1}$, where $L_{3,i}=L_{1}$ and $L_{3,ijk}=L_{3}$ are used as an approximation.
These time scales can be evaluated using numerical integration for the ground state density distribution obtained from solving Eq. (\ref{eq: GPE}).
Here, we further show the results for Rubidium atoms with non-zero two-body loss rate in Fig. \ref{cap: Rb-loss}.
The loss due to two-body effects is significantly larger than that due to three-body effects in this case, which limits the maximum cat size to $\bar{n}=10$ atoms, as compared to a few hundred atoms for the sodium example used
in the main text. Note that this is only one possible choice of states for Rubidium. Large cats may still be possible if appropriate internal states and other conditions can be found such that two-body loss is suppressed. 

For cat state creation, the maximum loss of atoms in the component $B$ cannot be larger than one atom.
Therefore, we are trying to give a conservative estimation.
Since $|\psi_{a}(r)|^2\le|\psi_{a0}(0)|^2=\mu_{a}/(N_{a}U_{aa})$ in the TFA, we can use the maximum $|\psi_{a0}(0)|$ for the main BEC, and the Gaussian $\phi_{b0}(r)$ for component B in Eq. (\ref{eq: GPE-unpert-sol}):
\begin{eqnarray}
\int\rho_{b}d^{3}r & = & n\label{eq: density-b}\\
\int\rho_{b}^{2}d^{3}r & \approx & \int d^{3}r|\phi_{b0}|^{4}n^{2}=(\sqrt{2}s_{b})^{-3}n^{2}\label{eq: density-bb}\\
\int\rho_{a}\rho_{b}d^{3}r & \approx & \left(\frac{\mu_{a0}}{U_{aa}}\right)n=\frac{15^{2/5}\pi^{1/5}}{8}\frac{N^{2/5}}{a_{a}^{3/5}s_{a}^{12/5}}n\label{eq: density-ba}\\
\int\rho_{b}^{3}d^{3}r & \approx & \int d^{3}r|\phi_{b0}|^{6}n^{3}=3^{-3/2}s_{b}^{-6}n^{3}\label{eq: density-bbb}\\
\int\rho_{a}\rho_{b}^{2}d^{3}r & \approx & \left(\frac{\mu_{a0}}{U_{aa}}\right)\int d^{3}r|\phi_{b0}|^{4}n^{2}=\frac{15^{2/5}\pi^{1/5}}{16\sqrt{2}}\frac{N^{2/5}}{a_{aa}^{3/5}s_{a}^{12/5}s_{b}^{3}}n^{2}\label{eq: density-bba}\\
\int\rho_{a}^{2}\rho_{b}d^{3}r & \approx & \left(\frac{\mu_{a0}}{U_{aa}}\right)^{2}n=\frac{15^{4/5}\pi^{2/5}}{64}\frac{N^{4/5}}{a_{aa}^{6/5}s_{a}^{24/5}}n\label{eq: density-baa}
\end{eqnarray}
The estimations for three body loss are shown in Fig. \ref{cap: loss-approx}, which suggests they are good lower bounds for $\tau_{3,ijk}$ and the time to lose one atom through all loss channels $\tau_\ell = (\tau_{1,b}^{-1} + \tau_{3,baa}^{-1} + \tau_{3,bba}^{-1} + \tau_{3,bbb}^{-1})^{-1}$.
The estimation is better at small $\omega_b$, since the density of component A is not repelled away so that the approximation $|\psi_{a}(0)|^2\approx |\psi_{a0}(0)|^2$ is good.

\section{Readout loss}

The readout loss from spin states to light is treated using the beam
splitter model with a given loss rate $\mathfrak{r}^{2}$. The state
passing through the beam splitter is $\left|\chi_{out}\right\rangle _{L}=\sum_{k=0}^{n}B_{nk}\left|n-k,k\right\rangle _{L}$
with $B_{nk}=\mathfrak{t}^{n-k}\mathfrak{r}^{k}n!/(k!(n-k)!)$, so
the reduced density matrix $\hat{\rho}'$ is
\begin{equation}
\hat{\rho}'=Tr_{2}(\hat{\rho})=\sum_{i}\left\langle i|\psi_{out}\right\rangle _{L}\left\langle \psi_{out}|i\right\rangle =\sum_{n,m}\sum_{k}^{\min(m,n)}B_{nk}B_{mk}^{*}c_{n}(t)c_{m}^{*}(t)\left|n-k\right\rangle \left\langle m-k\right|
\end{equation}
Hence, the resulting Q function with loss $Q_{loss}(s,\theta,t)$ and initial coherent state $|\alpha\rangle_L$
can be written as:
\begin{equation}
Q_{loss}(s,\theta,t)=\frac{1}{\pi}e^{-(t\alpha-s)^{2}}\sum_{m,n}\left(\sum_{k=0}^{\min(m,n)}\frac{(\alpha^{2}\mathfrak{r}^{2})^{k}(\mathfrak{t}\alpha s)^{n-k}(\mathfrak{t}\alpha s)^{m-k}}{k!(n-k)!(m-k)!}e^{-(\alpha^{2}\mathfrak{r}^{2}+2\mathfrak{t}\alpha s)}\right)e^{-i(n-m)\theta}e^{-i\left(\eta(n)-\eta(m)\right)t}
\end{equation}
The term inside the big bracket is the bivariate Poisson distribution
so this summation is upper bounded by 1. Therefore the resulting $Q_{loss}(\beta)$
is confined to the annulus $|s-\mathfrak{t}\alpha|\sim1$.
Hence, the effect of photon loss is to move the peak of the Q-function toward the origin, as shown in Fig. 3c and 3d in the main text.
This form of the Q-function can be evaluated fairly efficiently with time complexity
of order $\mathcal{O}(\bar{n}^{3/2})$, which allows us to evaluate it for cat sizes of order a few hundred atoms.

\section{Allowable uncertainty in atom number}

\begin{figure}
    \centering
    \subfigure[$d\eta_1/dN$ vs $\omega_{b}$]{\includegraphics[width=0.44\textwidth]{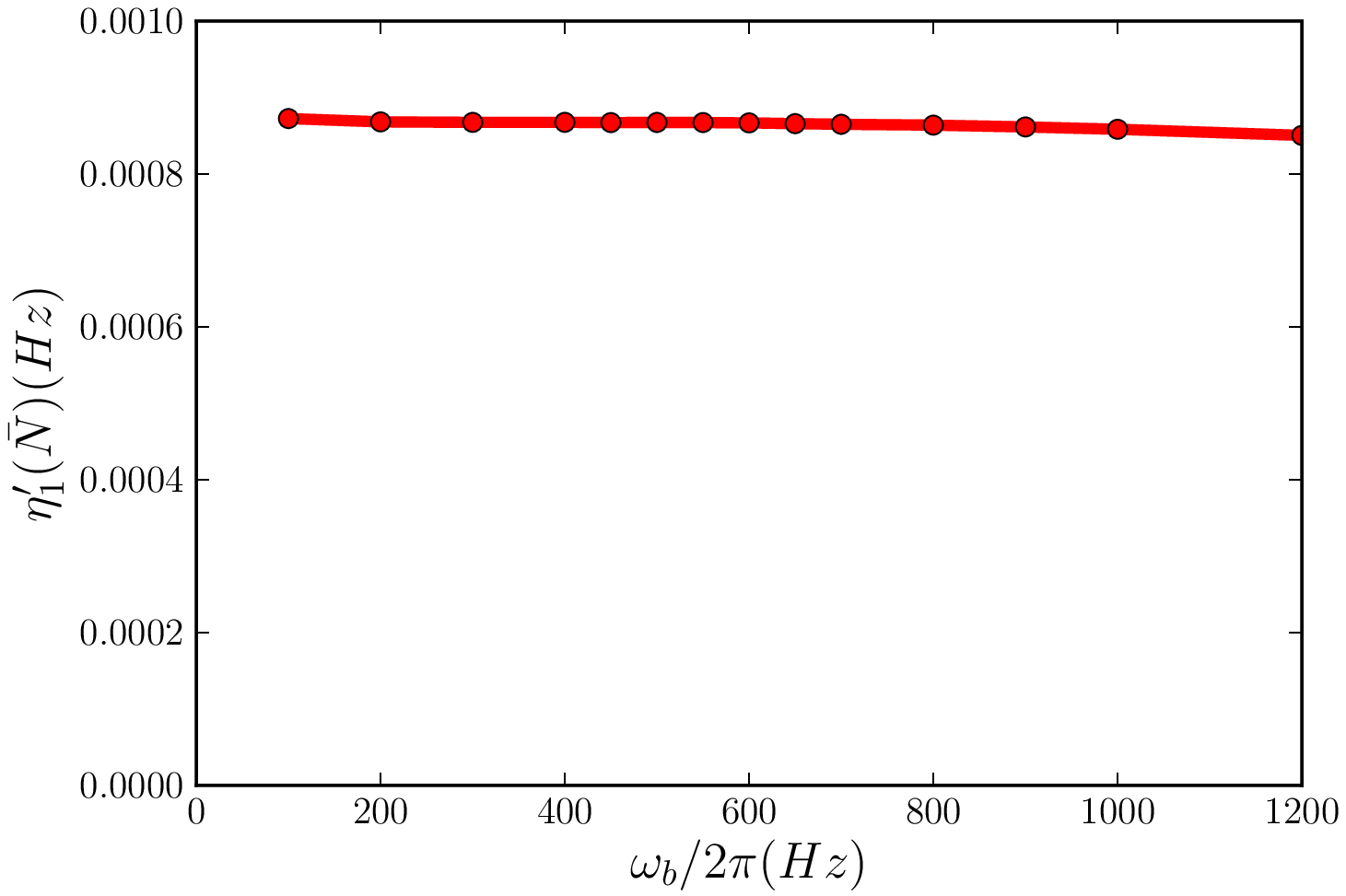}}
    \subfigure[$d\eta_2/dN$ vs $\omega_{b}$]{\includegraphics[width=0.44\textwidth]{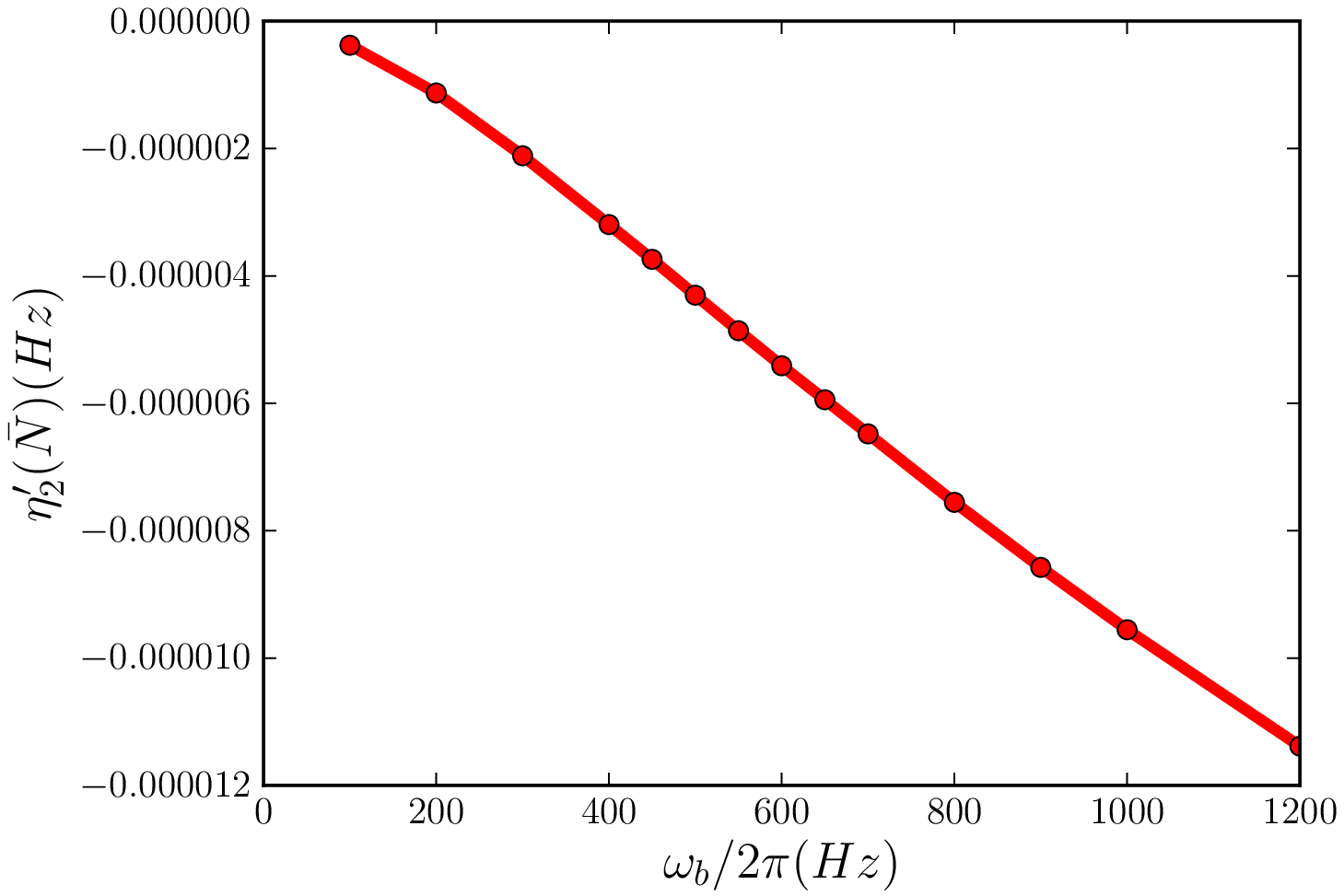}}
    \vfill
    \subfigure[Dephasing $\Delta\varphi$ with $\Delta N=0.05N$, at $\tau_c$]{\includegraphics[width=0.44\textwidth]{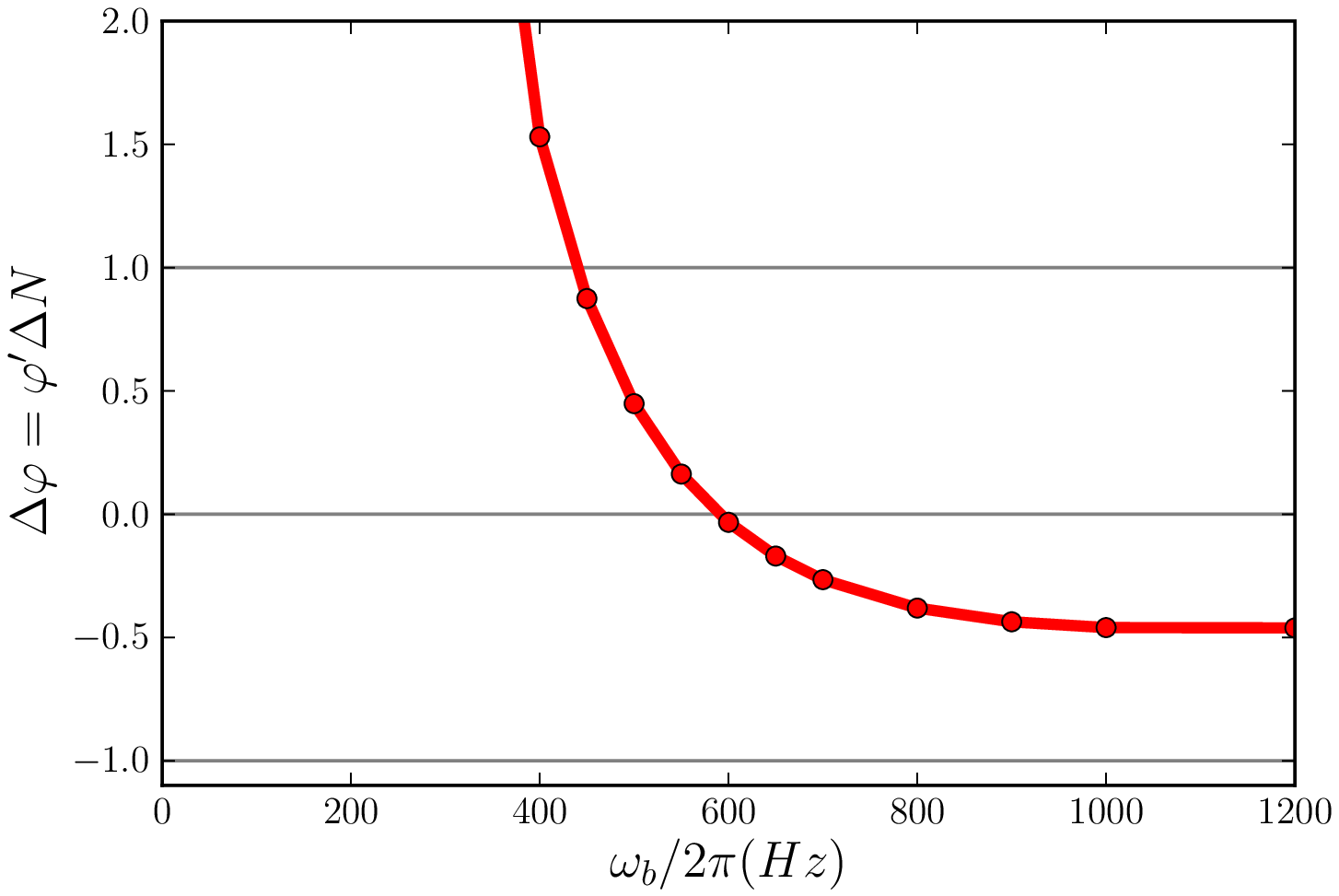}}
    \subfigure[Allowable range of uncertainty in $\Delta N/N$, at $\tau_c$]{\includegraphics[width=0.44\textwidth]{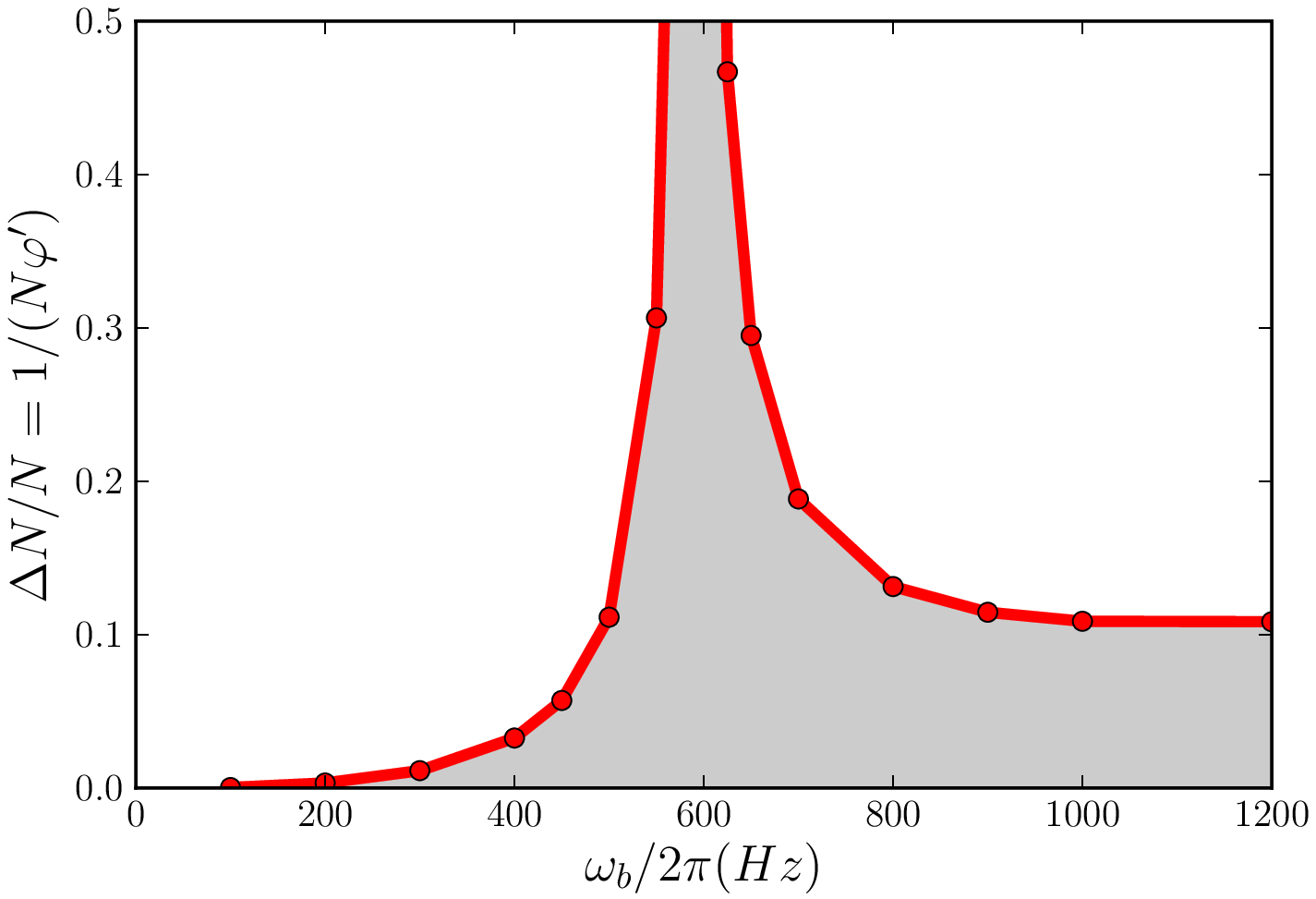}}
\caption{\label{cap: N-uncertainty}
Allowable range of atom number uncertainty with $\bar{N}=10^{5}$ and $\bar{n}=100$.
(a) $\eta_{1}'$ is basically constant over a large range of $\omega_{b}$.
(b) $\eta_{2}'$ decreases with $\omega_{b}$.
(c) Total dephasing $\Delta\varphi = \varphi'\Delta N$, where $\varphi'=(\pi/2\eta_{2})(\sum_{k}k\bar{n}^{k-1}\eta_{k}')$ includes up to fourth order terms from the GPE.
$\Delta\varphi$ needs to be smaller than 1 to have two distinguishable peaks of cat states, see Fig 3c and 3d in main text for the Q-function for the case of $\Delta \varphi=0.5$ at $\omega_b/2\pi = 500 \mbox{Hz}$.
(d) The gray region indicates the allowable uncertainty in atom number $\Delta N/N$.
It shows that the uncertainty in $N$ can be very large around $\omega_{b}/2\pi=600$, and about $10\%$ for high $\omega_{b}$.
The other parameters used are the same as in Fig. \ref{cap: Properties}.
}
\end{figure}

Since all $\eta_k(N)$ depend on the total atom number $N$, the statistical fluctuations in $N$ can cause dephasing (equivalent to angular spreading in phase space for the Q-function studied here), which can wash out all observable features of cat states (consider for example the $N$-dependent rotation $e^{-i\eta_{1}(N)t}$ caused by $\eta_1(N)$). The dephasing is small if the derivatives of the coefficients with respect to $N$, $\eta_k'(N)=\partial_{N}\eta_{k}(N)$, are small.
These quantities are plotted in Fig. \ref{cap: N-uncertainty}a and \ref{cap: N-uncertainty}b.
Note that the constancy of $\eta_1'$ in Fig. (\ref{cap: N-uncertainty}a) can be understood from Eq. (\ref{eq: eta1-approx}) because $\eta_1'=(U_{ab}/U_{aa}-1)\mu_{a0}'$ is independent of $\omega_b$.
Also, the dephasing is linear in time, hence, a short cat time $\tau_c$ can significantly reduce the dephasing effects.
 Moreover the rotation generated by $\eta_1(N)$ can be canceled by the opposite rotation generated by $\eta_2(N)$, as we derive below.

First considering the expansion of $n=\bar{n}+\Delta n$ around $\bar{n}$ the relevant terms become
\begin{equation}
\eta_{1}(N)n+\eta_{2}(N)n^{2}=(\eta_{1}\bar{n}+\eta_{2}\bar{n}^{2})+(\eta_{1}+2\eta_{2}\bar{n})\Delta n+\eta_{2}\Delta n^{2} \label{eq: etas_expand_n}
\end{equation}
On the right hand side, the first term gives a global phase which can be neglected. The second term leads to a rotation in phase space. Writing $N=\bar{N}+\Delta N$ and expanding the coefficients around $\bar{N}$ one has
\begin{equation}
\eta_{k}(N)=\eta_{k}(\bar{N)}+\eta_{k}'(\bar{N})\Delta N
\end{equation}
where $\eta_{k}'(N)=\partial_{N}\eta_{k}(N)$. Note that $\Delta\eta_{k}(N)=\eta_{k}'(N)\Delta N$ is the fluctuation in $\eta_{k}$ due to the uncertainty $\Delta N$. Substituting these back into the second term in Eq. (\ref{eq: etas_expand_n}) yields the dephasing term $\left(\eta_{1}'\Delta N+2\bar{n}\eta_{2}'\Delta N\right)\Delta n$. This dephasing term is the source of a $\Delta N$ dependent rotation in the $\beta$-plane, which is eliminated when the condition $\eta_{1}'(\bar{N})+2\bar{n}\eta_{2}'(\bar{N})=0$ is satisfied, see Fig. \ref{cap: N-uncertainty}c. In particular, we want to find out the maximum allowable $\Delta N$ that still preserves an observable spin cat state at the cat time $\tau_{c}$. Therefore, we define $\Delta\varphi=\varphi'\Delta N=\tau_{c}(\eta_{1}'+2\bar{n}\eta_{2}')\Delta N$,
and the condition $|\Delta\varphi|\lesssim1$ should be satisfied, or
\begin{equation}
\Delta N\lesssim\frac{1}{\varphi'}
\end{equation}
The higher order terms $\eta_{k}$ can also be included, yielding
\begin{equation}
\varphi'=\frac{\pi}{2\eta_{2}}(\sum_{k}k\bar{n}^{k-1}\eta_{k}')
\end{equation}
Numerically, we find $\eta_k'(N)$ by taking the numerical derivative of $\eta_k(N)$.
The results in Fig. \ref{cap: N-uncertainty}d show that there is a large range of allowable uncertainty in atom number $\Delta N$ if $\omega_{b}$ is high enough.
Note that this range is an estimation since only the first order approximation of $\Delta\eta_{k}=\eta_{k}'(N)\Delta N$ is used.
In contrast, the accuracy requirement $\Delta N/N$ at low trapping $\omega_b$ is even higher than the high resolution of counting cold atoms of 1 in 1200 in a recent experiment \cite{S_hume_accurate_2013}.

\section{Atom loss}

\begin{figure}
    \centering
    \subfigure[$t=\tau_{c},L_{1}\tau_{c}=0.01$]{\includegraphics[width=0.45\textwidth]{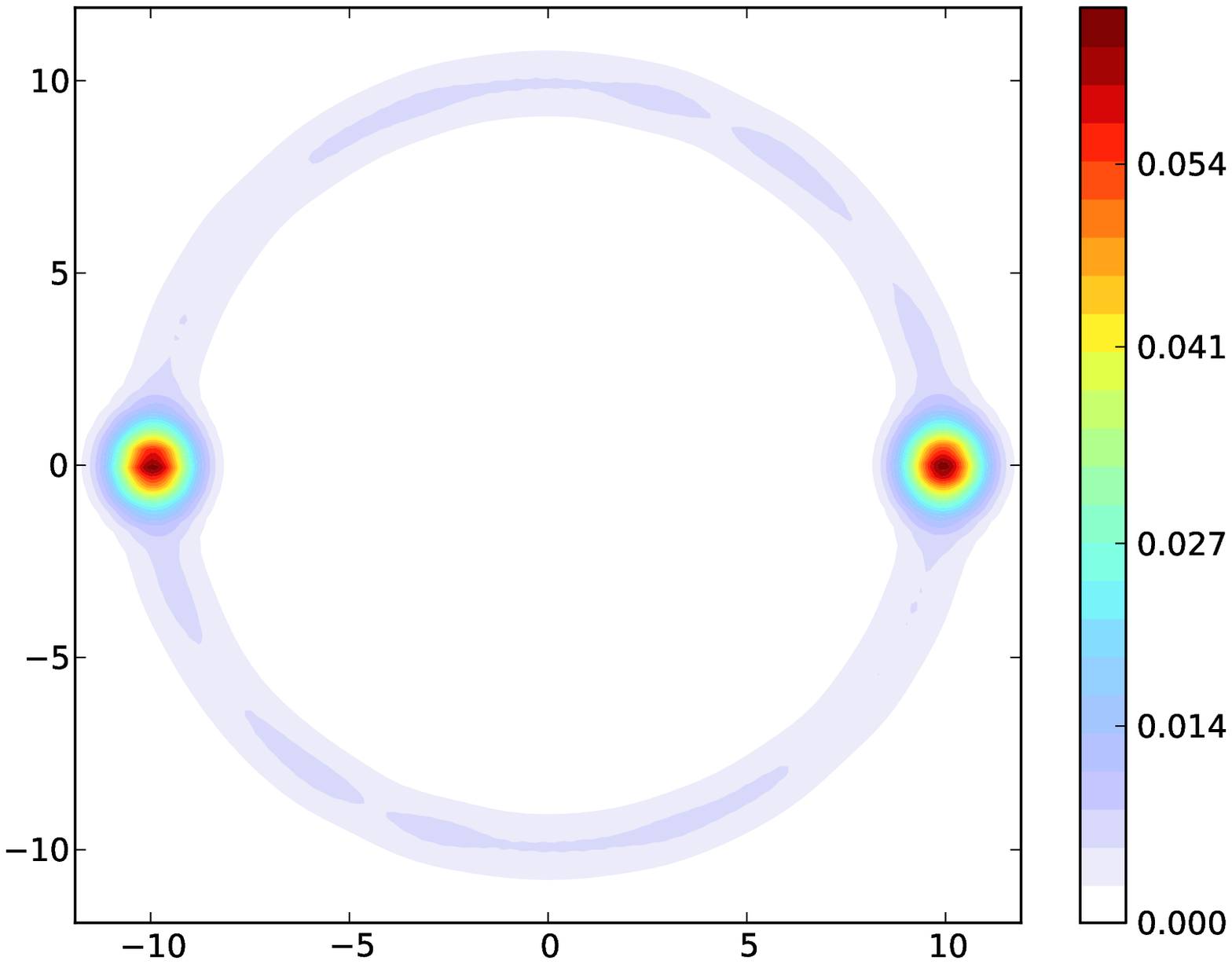}}
    \subfigure[$t=2\tau_{c},L_{1}\tau_{c}=0.01$]{\includegraphics[width=0.45\textwidth]{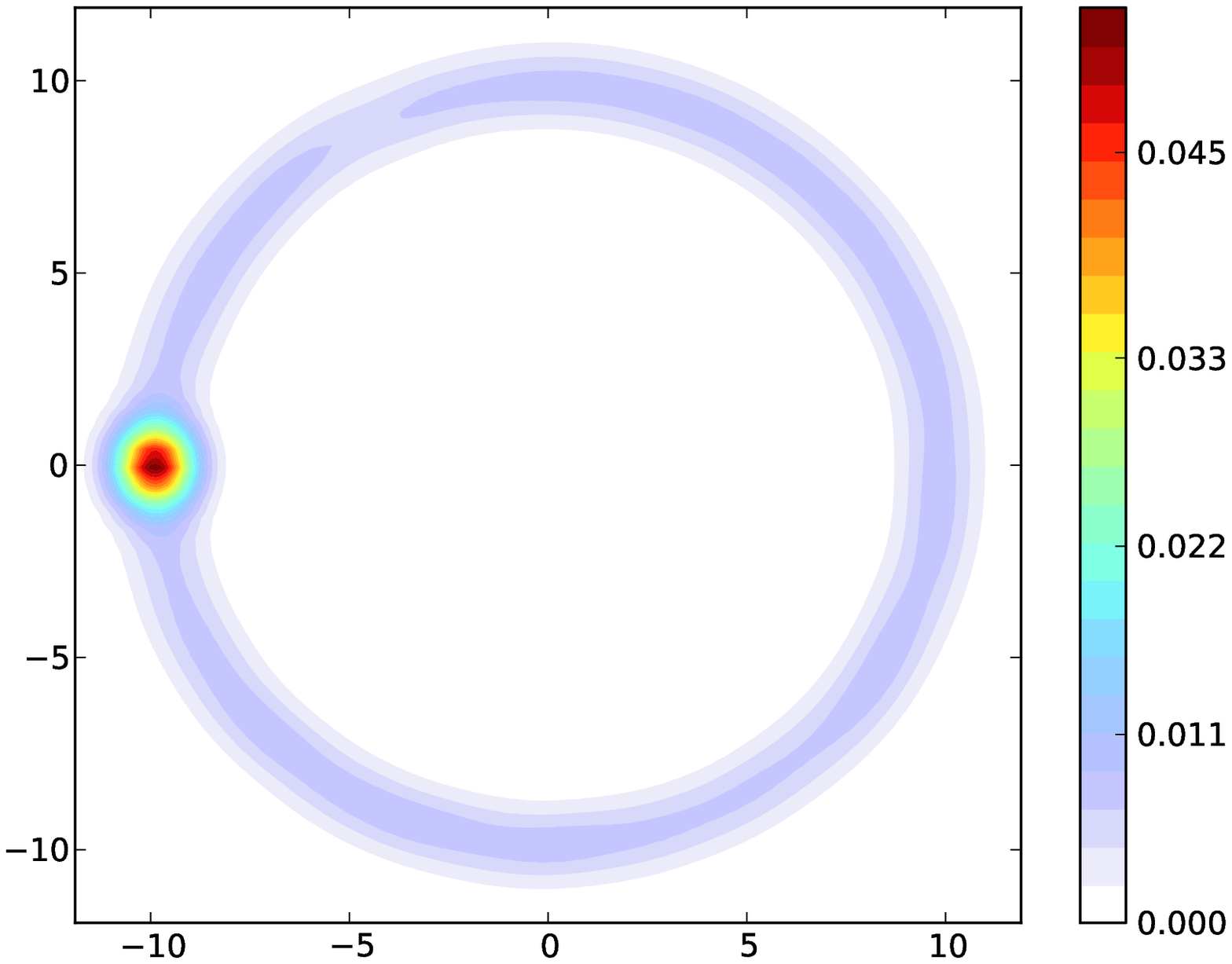}}
    \vfill
    \subfigure[$t=\tau_{c},L_{1}\tau_{c}=0.025$]{\includegraphics[width=0.45\textwidth]{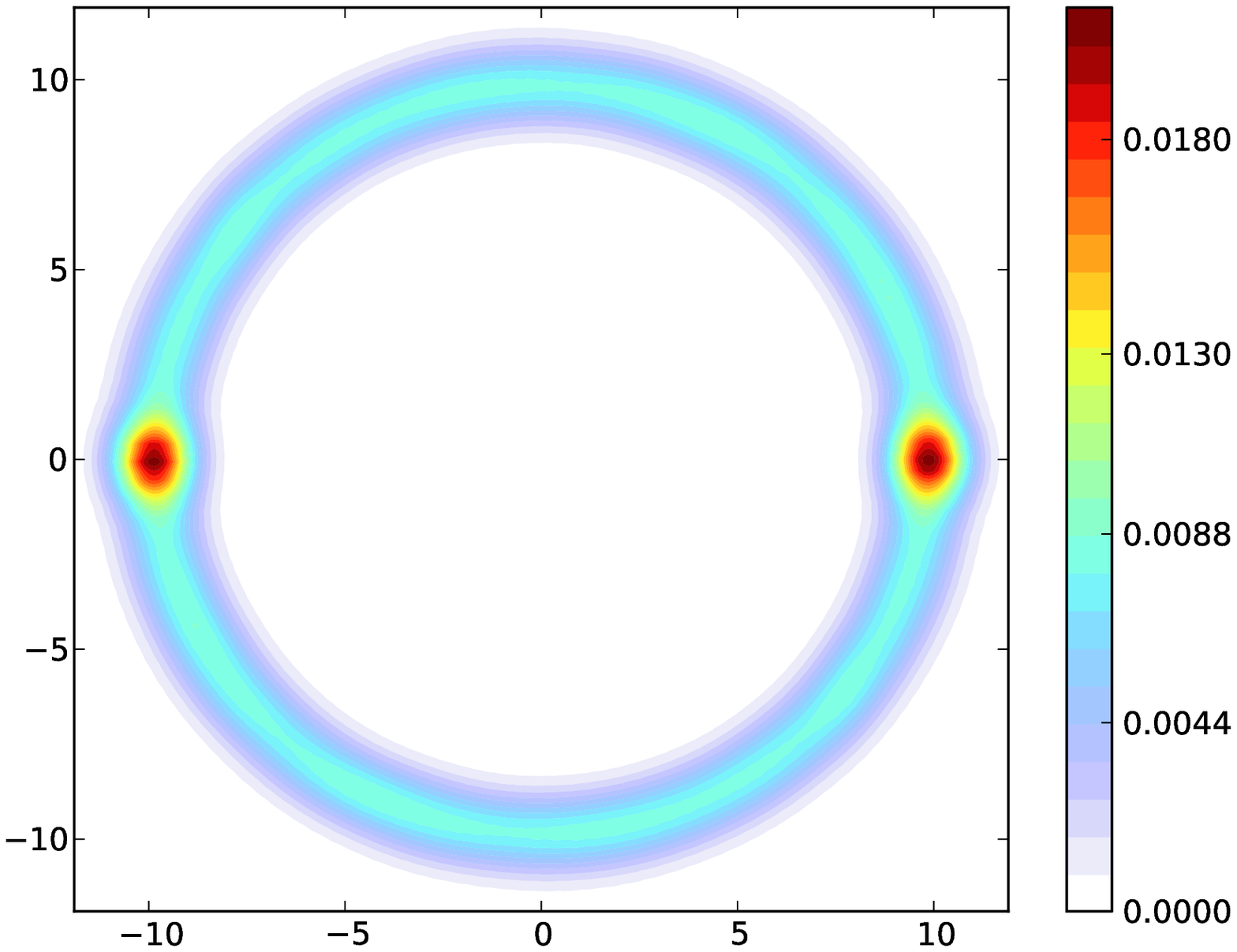}}
    \subfigure[$t=2\tau_{c},L_{1}\tau_{c}=0.025$]{\includegraphics[width=0.45\textwidth]{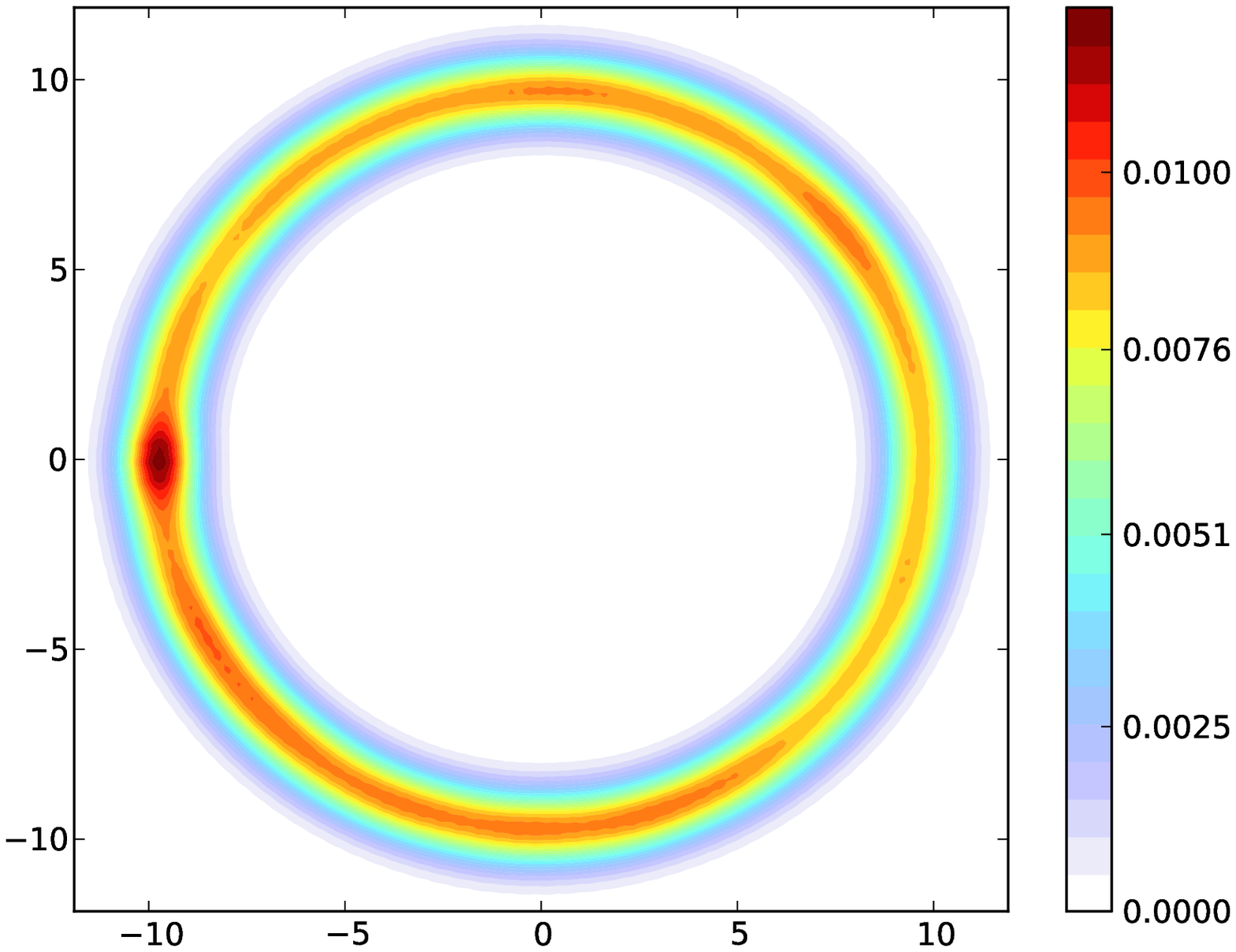}}
\caption{\label{cap: Q-loss-standard}
Q function of continuous atom loss for the standard Kerr effect with Hamiltonian $\hat{\mathcal{H}}=\hbar\eta_{2}\hat{n}^{2}$.
(left column) At cat time $t=\tau_{c}=\pi/|2\eta_{2}|$,
(right column) At revival time $t=2\tau_{c}$. Mean photon number $\bar{n}=100$ and 5000 samples.
}
\end{figure}

\begin{figure}
    \centering
    \subfigure[]{\includegraphics[width=0.45\textwidth]{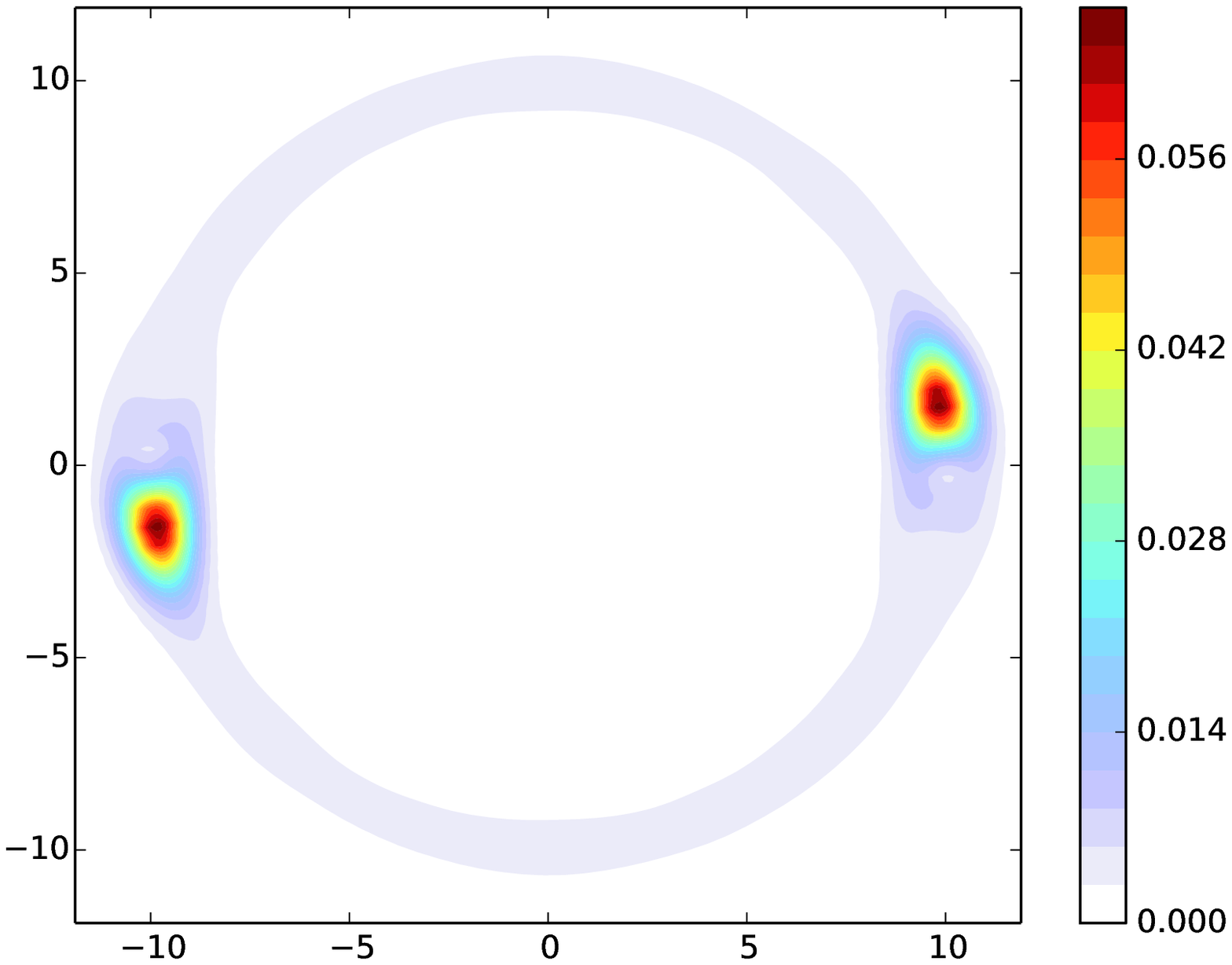}}
    \subfigure[]{\includegraphics[width=0.45\textwidth]{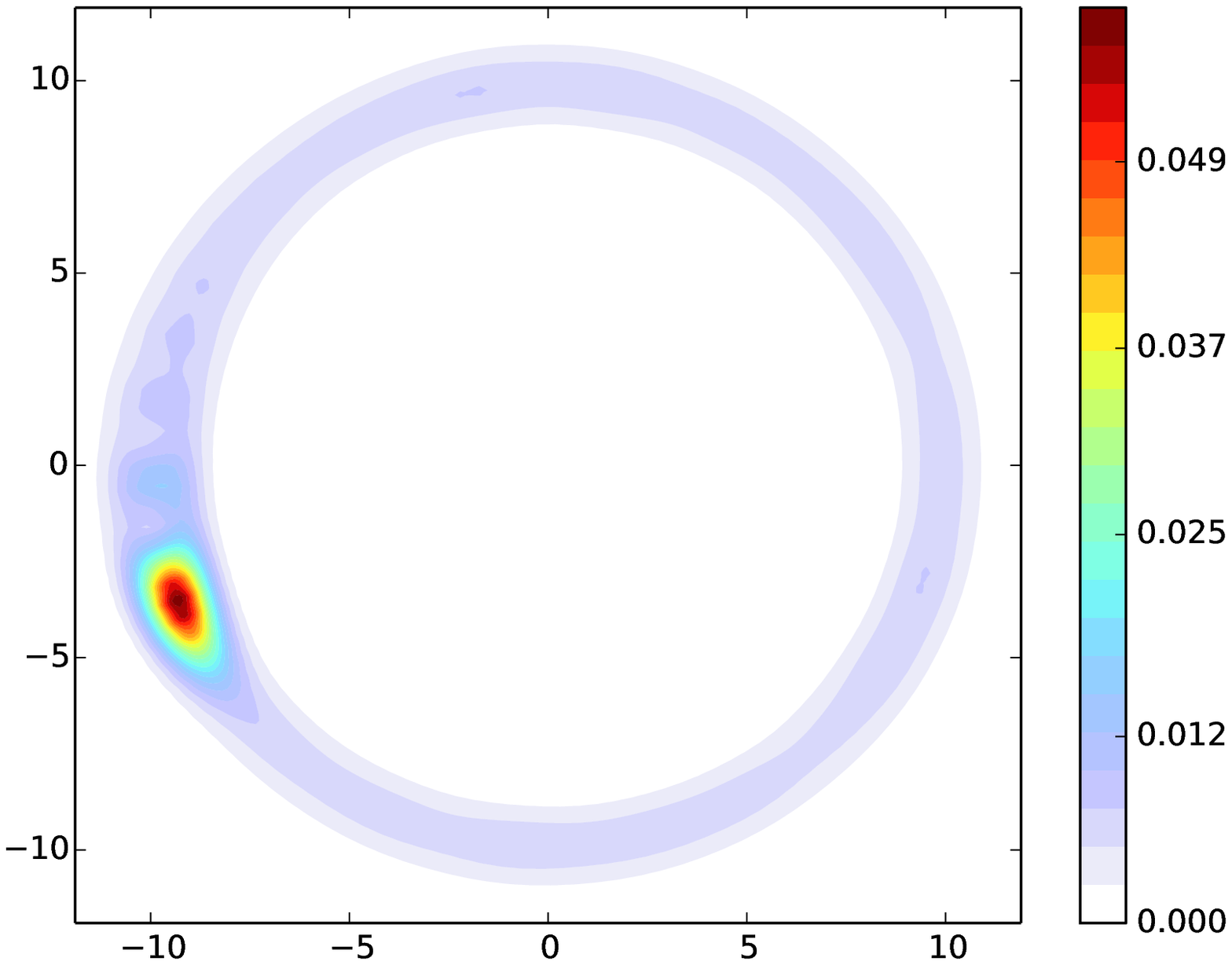}}
\caption{\label{cap: Q-loss-BEC-real-standard}
Q function of continuous atom loss for $\eta_{k}$ calculated from GPE for the parameters given in Fig. 2b in the main text ($\bar{n}=100$, $\omega_{b}=2\pi500$).
(a) $t=\tau_{c}^{*}=1.06\tau_{c}=0.68s$, $L_{1}=0.01/s$ corresponding to about 0.68 atom loss,
(b) $t=2\tau_{c}^{*}$, about 1.34 atom loss.
No other effect is included. 5000 samples.
}
\end{figure}

The continuous loss of atoms from the BEC can be described by the master equation:

\begin{equation}
\dot{\rho}=-\frac{i}{\hbar}[\hat{\mathcal{H}},\rho]+\sum_{i}(\hat{R}_{i}\rho\hat{R}_{i}^{\dagger}-\frac{1}{2}\hat{R}_{i}^{\dagger}\hat{R}_{i}\rho-\frac{1}{2}\rho\hat{R}_{i}^{\dagger}\hat{R}_{i})
\end{equation}
where $\rho$ is the density operator,
$\hat{\mathcal{H}}$ is the Hamiltonian of the system,
$\hat{R}$ is the Lindblad operator for the loss channel in question, and the summation is over the different loss channels.
In the main text, the 2-body loss is assumed to be zero and 3-body loss is much lower than the
1-body loss (see Fig. 3) in the regime considered. Therefore, only 1-body
loss will be considered below to simplify the calculation.
The loss atoms from the large component causes a change in $N$. The resulting effects are similar to the fluctuations in $N$ considered in the previous section. Here we therefore focus on the small component. In this case one can effectively describe the system by the density operator $\rho=|\chi(t)\rangle \langle \chi(t)|$, with the initial state $|\chi(t=0)\rangle=|\alpha\rangle=\sum c_{n}|n\rangle$. Only one Lindblad operator $\hat{R}=\sqrt{L_{1}}\hat{a}$
is needed. The simplified master equation is:
\begin{equation}
\dot{\rho}=-\frac{i}{\hbar}[\hat{\mathcal{H}},\rho]+L_{1}(\hat{a}\rho\hat{a}^{\dagger}-\frac{1}{2}\hat{a}^{\dagger}\hat{a}\rho-\frac{1}{2}\rho\hat{a}^{\dagger}\hat{a})
\end{equation}
with the Hamiltonian given by Eq. (\ref{eq: etas}). The method used to simulate
the system is the Quantum Jump Method \cite{S_Garraway94,S_Plenio-Knight98}.
The results are shown in Fig. \ref{cap: Q-loss-standard}.
For the standard Kerr effect without higher-order terms,
it can be observed that the creation of spin cat state still results in two clear peaks in the Q-function even when 2.5 atoms are lost on average. In fact, the cat state is still visible even for an average loss of 5 atoms. However, for the detection scheme, the system is required to evolve for $2\tau_{c}$, which limits the loss rate to $L_{1}\tau_{c}<0.025$ as shown in the figure.
For the parameters used in Fig. 2b in the main text (including higher-order nonlinearities), the average number of atoms lost is only 0.68 and the effect of the loss is small, see Fig. \ref{cap: Q-loss-BEC-real-standard}. The main effect of the loss is a fairly uniform background ring in the Q function.

\section{Comparison with photon-photon gate proposal}

Ref. \cite{S_Rispe} utilizes a similar collision induced cross-Kerr nonlinearity in BEC to implement photon-photon gates, while the current scheme uses a self-Kerr nonlinearity to create spin cat states.
The Kerr effect in the previous scheme is enhanced by increasing both scattering length (through a Feshbach resonance) and the trapping frequency for both components.
However, the Feshbach resonance induced atom loss can be very large \cite{S_chin_feshbach_2010}, which will limit the maximum cat size.
Not relying on a Feshbach resonance also makes it possible to use the magnetic field to further eliminate atom loss.
Also, both trapping frequencies should not be increased at the same time because it will result in high atom loss through the collision with the main BEC.
Instead, we suggest here to increase only the trapping frequency of small component.
This results in a similarly strong Kerr effect, but with lower atom loss.

Moreover, the treatment in the previous paper, which used the quantized mean-field GPE with TFA and first order perturbation theory, does not allow the study of higher order nonlinearities or atom number fluctuations. Our present approach allows us to study both of these effects, and we show that they can be significant.
The assumption of equal trapping frequencies also limits the previous treatment to the non-phase separated regime, which limits the choice of regimes with low atom loss, such as the sodium atom example used here.
Furthermore, the density of the stored component in the previous scheme is much smaller (about four orders of magnitude) than the main component.
This raises the concern of other possible dominant effects on the same scale, such as quantum depletion \cite{S_Trail}.
As we have shown, these problem can be minimized in the current scheme by using a high enough $\omega_b$ so that the small component is located at the center with high density.
Note that Eq. (\ref{eq: eta2-approx-simple}), which is obtained as a limiting case for high $\omega_b$ here, is basically equivalent to the results of the treatment in Ref. \cite{S_Rispe}.

\end{document}